%% file: main.tex
\documentclass{article} %
\usepackage{iclr2027_conference,times}

\input{math_commands.tex}

\usepackage[utf8]{inputenc} %
\usepackage[T1]{fontenc}    %
\PassOptionsToPackage{hyphens}{url} %
\usepackage{hyperref}       %
\usepackage{url}            %
\usepackage{longtable}
\newlength{\benchmarkmodelwidth}
\usepackage{booktabs}       %
\usepackage{nicefrac}       %
\usepackage{microtype}      %
\usepackage[table]{xcolor}  %

\usepackage{amsmath,amsfonts}
\usepackage{amssymb}
\usepackage{amsthm}
\usepackage{mathtools}
\usepackage{lettrine}
\usepackage{multirow}
\usepackage{multicol}
\usepackage{tablefootnote}
\usepackage{enumerate}
\usepackage{enumitem}
\usepackage{extarrows}
\usepackage{caption}
\usepackage{graphicx}
\usepackage{listings}
\usepackage{textcomp}
\usepackage{pifont}
\usepackage{marvosym}
\usepackage{makecell}
\usepackage{mdframed}
\usepackage{wrapfig}
\usepackage{needspace}
\usepackage{placeins}
\usepackage{tcolorbox}
\tcbuselibrary{listings, skins, breakable}  %

\newtcblisting[auto counter]{paperprompt}[2]{
  enhanced,
  breakable=false,
  listing only,
  colback=white,
  colframe=gray!60,
  colbacktitle=gray!10,
  coltitle=black,
  fonttitle=\normalfont\bfseries,
  title={Prompt \thetcbcounter: #1},
  label={#2},
  boxrule=0.4pt,
  arc=0pt,
  left=8pt,right=8pt,top=6pt,bottom=6pt,
  before skip=10pt,after skip=10pt,
  listing options={
    basicstyle=\normalfont\fontsize{10}{12}\linespread{1}\selectfont\color{black},
    upquote=true,
    breaklines=true,
    breakatwhitespace=true,
    breakindent=0pt,
    breakautoindent=false,
    moredelim={[is][\bfseries]{**}{**}},
    columns=fullflexible,
    keepspaces=true,
    showstringspaces=false,
    backgroundcolor=\color{white},
    frame=none,
    aboveskip=0pt,belowskip=0pt
  }
}

\theoremstyle{definition}

\definecolor{baselinegray}{RGB}{240, 240, 240}  %
\definecolor{mymodelblue}{RGB}{101, 189, 186}   %

\definecolor{mymodelgreen}{RGB}{225, 240, 225}   %

\definecolor{improvementblue}{cmyk}{0.1, 0, 0, 0}
\definecolor{textgray}{gray}{0.55} 
\definecolor{deltablue}{cmyk}{1, 0, 0, 0} %
\definecolor{tboxblue}{RGB}{60, 155, 201}
\definecolor{palettepink}{RGB}{252, 117, 123}        %
\definecolor{palettecoral}{RGB}{249, 127, 95}        %
\definecolor{palettepeach}{RGB}{250, 162, 111}       %
\definecolor{palettesand}{RGB}{253, 205, 148}        %
\definecolor{palettecream}{RGB}{254, 225, 153}       %
\definecolor{palettegreen}{RGB}{176, 214, 169}       %
\definecolor{paletteteal}{RGB}{101, 189, 186}        %
\definecolor{paletteblue}{RGB}{60, 155, 201}         %
\definecolor{classcolor}{HTML}{FCE5CD}
\definecolor{vqacolor}{HTML}{D9E1F2}
\definecolor{retcolor}{HTML}{E2EFDA}
\definecolor{grdcolor}{HTML}{F2DCDB}
\definecolor{finalcolor}{HTML}{DEEBF6}
\definecolor{oodcolor}{HTML}{FFF2CC}

\title{Reasoning Quality Matters: Combating Reasoning Collapse in LLM-based Embedding Learning}

\input{authors.tex}

\iclrfinalcopy
\hypersetup{
  hidelinks,
  pdftitle={Reasoning Quality Matters: Combating Reasoning Collapse in LLM-based Embedding Learning},
  pdfauthor={Zihan Gong, Xiaohan Ye, Jiangchao Yao, Jinsong Lan, Xiaoyong Zhu, Xu Chen}
}

\begin{document}

\maketitle
\lhead{Preprint}

\begin{abstract}
Large Language Models (LLMs) have recently shown strong potential for producing context-rich text embeddings for retrieval. Most existing methods either treat embedding learning as passive feature extraction or exploit LLM reasoning through instruction following for better embedding optimization. However, specialization toward embedding objectives can suppress useful reasoning generation or produce retrieval-irrelevant text. We refer to these two forms of degradation as \emph{reasoning collapse}.
To address this issue, we propose \textbf{CoFree} (\textbf{Co}llapse-\textbf{Free} Reasoning Embedding), a two-stage framework that progressively integrates LLM reasoning into query and document embedding optimization while preserving reasoning quality. At the first stage, CoFree applies reference-guided supervised fine-tuning to restore the reasoning ability and retain representational strength of the foundation embedding model. At the second stage, we introduce dual rewards, an embedding-oriented reward and a reasoning-oriented reward, to guarantee fine-grained reasoning of the relevance toward the embedding goal in reinforcement learning. This endpoint-coupled optimization transforms embedding learning from static alignment into a high-quality reasoning-guided search process for retrieval. Extensive experiments demonstrate the effectiveness of CoFree, with CoFree-4B achieving an average absolute improvement of \textbf{2.8} nDCG@10 points over Qwen3-Embedding-4B across 22 datasets from MTEB and BRIGHT. Online experiments in a real-world retrieval system further show consistent gains. Code, RTED, and model checkpoints will be made publicly available.
\end{abstract}

\section{Introduction}
Dense text embeddings are a fundamental component of modern information retrieval. Historically, language modeling and embedding learning have been developed as separate paradigms: generative models are optimized to produce coherent text~\citep{touvron2023llama,bai2023qwen,abdin2024phi}, whereas embedding models compress text into discriminative semantic representations~\citep{mikolov2013efficient,devlin2019bert,raffel2020exploring}. Large language models (LLMs) create an opportunity to connect these capabilities: before producing a vector, a model can explicitly analyze the intent, concepts, and relations expressed by an input, potentially exposing fine-grained and compositional semantics that are difficult to capture through direct compression alone. 

Recent methods have therefore adapted generative LLMs into strong text encoders through instruction-based representation learning~\citep{behnamghader2024llm2vec,chen2024bge,zhang2025qwen3}. Some methods focus on domain and language generalization~\citep{li2023towards,zhang2024mgte,xiao2024c,lee2024nv}. For instance, MGTE~\mbox{\citep{zhang2024mgte}} targets long-context and multilingual retrieval scenarios, producing embeddings robust to large discourse spans and cross-lingual variance.
To further exploit the reasoning abilities of LLMs, a newer line of work generates additional text from a query or document and appends it to the original input for embedding extraction. This additional generated text is referred to as \emph{reasoning text}. For example, TTE~\citep{cui2025think} uses chain-of-thought-style reasoning, while LREM~\citep{tang2026large} uses keyword-style reasoning. Search-R3~\citep{gui2025search} and UME-R1~\citep{lan2025ume} apply reinforcement learning to explore retrieval-oriented reasoning trajectories.

However, generating reasoning does not guarantee that the reasoning remains useful for retrieval. We observe two complementary failure modes. First, specialization toward embedding objectives can suppress the model's generative capability, resulting in empty outputs, uncontrolled repetition, or incoherent text, and we call this \emph{generation collapse}. Second, even when generation remains fluent, supervision focused primarily on the final embedding can produce reasoning that is generic, irrelevant, or insufficiently discriminative, and we call this \emph{semantic collapse}. For example, reasoning that conflates the concept \textit{coral reefs} with the unrelated entity \textit{Coral Reef Casino} simply because they share surface tokens.
We refer to these failures collectively as \textbf{\textit{reasoning collapse}}. Consequently, naively injecting reasoning can be worse than using no reasoning at all (See Section~\ref{sec:reasoning_quality}). Collapsed reasoning either corrupts the generative pathway or injects misleading semantics into the embedding, so the central challenge is not whether to reason but how to keep reasoning retrieval-faithful.
This raises two questions: (1) \textit{How can an embedding model recover useful reasoning generation without sacrificing its pretrained representation quality?} (2) \textit{How can the restored reasoning be optimized to distinguish relevant documents from hard negatives, rather than merely remaining fluent?}

To address these questions, we propose \textbf{Collapse-Free Embedding Learning (CoFree)}, a two-stage framework that builds on an existing LLM-based embedding model and progressively addresses the two forms of reasoning collapse. The first stage, \emph{Reasoning Restoration Learning}, uses supervised fine-tuning (SFT) to recover reasoning generation while preserving the backbone's embedding competence. It jointly optimizes a language modeling loss, a contrastive embedding loss, and a reference-guided loss that anchors reasoning-conditioned representations to those of the frozen backbone. The second stage, \emph{Reasoning Augmentation Reinforcement Learning} (RL), improves the retrieval utility of the restored reasoning through a dual-reward scheme. An \textbf{embedding-oriented reward} evaluates the positive--negative similarity margin of the final embeddings, while a \textbf{reasoning-oriented reward} uses a relevance reward model to assess how well the reasoning-augmented inputs distinguish relevant from irrelevant documents. The former provides endpoint supervision over the resulting vectors, whereas the latter provides pathway supervision over the reasoning. CoFree thereby optimizes both discriminative embeddings and retrieval-relevant reasoning trajectories. Beyond improving retrieval accuracy, our analyses in Section~\ref{sec:reasoning_quality} show that CoFree progressively mitigates reasoning collapse across SFT and RL, and that the resulting reasoning remains retrieval-useful when transferred to diverse frozen embedding backbones. Our contributions are summarized as follows:

\begin{itemize}
\vspace{-4pt}
    \item We identify and characterize \textit{reasoning collapse} in LLM-based embedding learning as two complementary failures: degradation of reasoning generation and degradation of the retrieval-relevant semantics carried by otherwise fluent reasoning. 
\vspace{-4pt}
    \item We propose CoFree, a two-stage framework that progressively mitigates reasoning collapse. Reference-guided SFT restores reasoning generation while retaining pretrained embedding competence, and dual-reward RL jointly optimizes embedding discrimination and the retrieval utility of generated reasoning.
\vspace{-4pt}
    \item We construct RTED, a 3.6M-instance resource for reasoning-based text embedding learning, covering diverse retrieval domains and produced through a fine-grained reasoning-data construction and filtering pipeline.
\vspace{-4pt}
    \item Extensive experiments on 22 datasets from MTEB and BRIGHT show that CoFree-4B improves over Qwen3-Embedding-4B by an average absolute margin of \textbf{2.8} nDCG@10 points. The effectiveness is also verified in a real-world information retrieval system.
\vspace{-10pt}
\end{itemize}

\section{Related Works}
\textbf{Era of Statistical and Deep Learning Methods.}
Early embedding methods such as TF-IDF~\citep{ramos2003using} and BM25~\citep{robertson2009probabilistic} represent documents as sparse term-weight vectors. While efficient, they fail to capture deeper semantics and are brittle to synonymy or conceptual variation.
The deep learning era brought dense contextualized representations. BERT~\citep{devlin2019bert} introduced bidirectional pretraining via masked language modeling. Sentence-BERT~\citep{reimers2019sentence} adapted this into a siamese architecture for direct sentence comparison. BGE-M3~\citep{chen2024bge} extended this paradigm to multi-lingual and multi-granular settings, while CLIP~\citep{radford2021learning} applied contrastive learning~\citep{oord2018representation,chen2020simple,zheng2021contrastive} to vision-language alignment.
Despite these advances, most approaches treat embedding learning as a fixed encoding process, without leveraging generation as an active optimization mechanism.

\textbf{Era of Large Language Models.}
The emergence of large language models (LLMs) has revolutionized embedding learning. A dominant trend remains contrastive learning, where embeddings are trained to pull similar pairs together and push unrelated ones apart. \citet{li2023towards} proposed staged optimization from coarse to fine-grained alignment. CCPairs~\citep{wang2022text} leveraged noisy aligned pairs for scalable embeddings, while MGTE~\citep{zhang2024mgte} extended this to multilingual and long-context settings, and C-Pack~\citep{xiao2024c} focused on Chinese embeddings.
Recent works have moved beyond static encoding. LLM2Vec~\citep{behnamghader2024llm2vec} and GRIT~\citep{muennighoff2024generative} showed that generative LLMs can serve as powerful encoders when strategically prompted. Qwen3-Embedding~\citep{zhang2025qwen3} integrated embedding generation with reranking in a unified foundation model. Search-R3~\citep{gui2025search} and UME-R1~\citep{lan2025ume} unify reasoning and embedding generation, producing search embeddings enriched with explicit reasoning. TTE~\citep{cui2025think} introduced an explicit thinking stage before embedding for vision-language contexts.
However, these works often overlook reasoning semantics, leading to reasoning collapse, where generated reasoning becomes semantically vacuous. We address this through a two-stage design: an SFT stage to safely restore reasoning capabilities, and an RL stage with a dual-reward scheme that validates semantic relevance. This ensures fine-grained, interpretable embeddings while mitigating collapse.

\section{Methodology}
The training pipeline, illustrated in Figure~\ref{fig:framework}, consists of two stages: (i) Reasoning Restoration Learning, which unlocks the model's generative and reasoning capabilities; and (ii) Reasoning Augmentation RL, which optimizes embeddings through high-quality explicit reasoning. Details are provided below.

\subsection{Reasoning Restoration Learning}
Following single-embedding task training, foundation models lose their text generation capabilities~\citep{behnamghader2024llm2vec}. We therefore apply reference-guided supervised fine-tuning (SFT) to reactivate their reasoning and generation abilities while maintaining the original embedding ability. To this end, we construct reasoning texts for queries and documents. These texts clarify the original semantics and provide relevant elaboration, bridging the semantic gap between inputs and targets. This activates the model's capacity for deeper semantic understanding and generation. The detailed construction pipeline is presented in Appendix~\ref{sec:data_pipeline}.

\begin{figure*}[t]
\centerline{\includegraphics[width=0.95\linewidth]{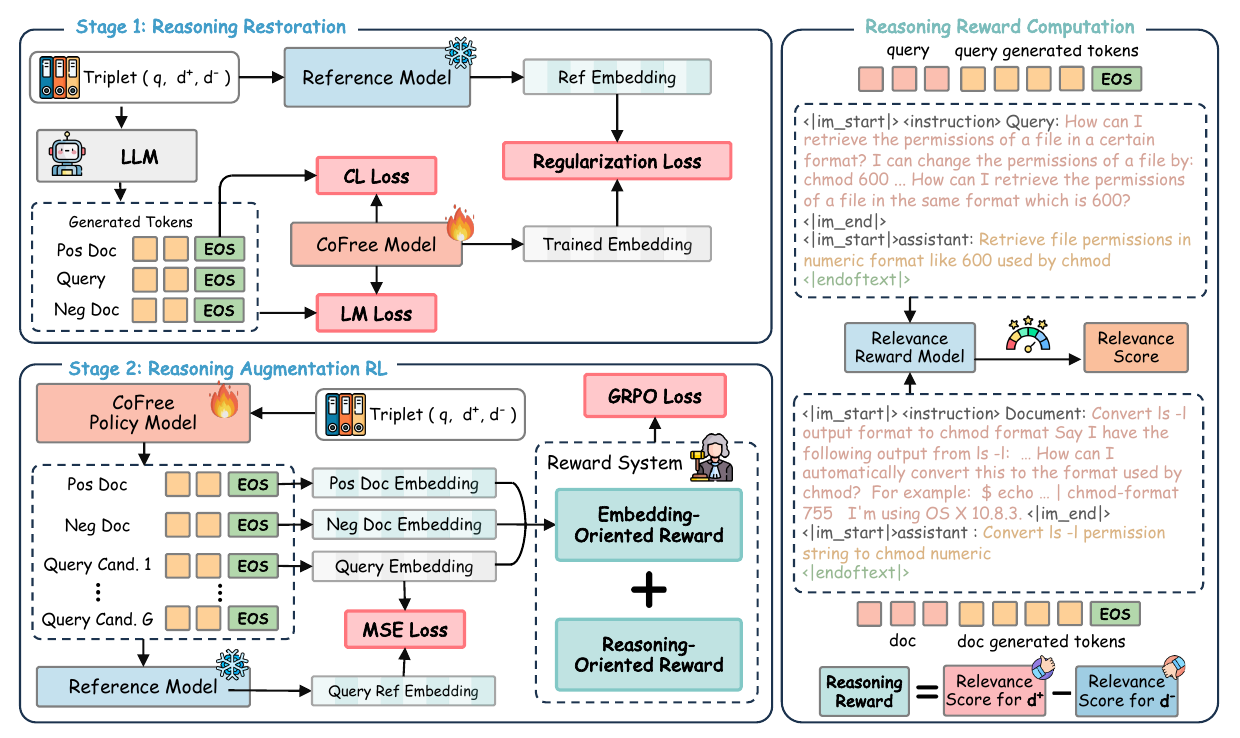}}
\caption{
The CoFree framework consists of two stages. In Stage 1, we jointly optimize contrastive loss, language modeling loss, and a reference-guided loss to safely restore the backbone's reasoning capability. In Stage 2, a dual-reward RL scheme evaluates both the semantic relevance of reasoning text and the embedding alignment quality.
}
\label{fig:framework}
\vspace{-15pt}
\end{figure*}

\textbf{Language Modeling and Contrastive Learning}:
As illustrated in the upper-left of Figure~\ref{fig:framework}, given an instruction along with a query or document as input, our model generates reasoning text followed by an \texttt{<eos>} token. 
We use the constructed texts as supervision targets and optimize the model with a standard causal language loss, as defined in Eq.~\eqref{eq:clm}.
\begin{equation}
\mathcal{L}_{\mathrm{LM}} = -\frac{1}{N} \sum_{i=1}^{N} \sum_{t=1}^{T_i} \log p_\theta\!\left(y_{i,t} \mid y_{i,<t}, x_i\right)
\label{eq:clm}
\end{equation}
where \(x_i\) denotes the \(i\)-th input sequence (instruction concatenated with the text), \(y_{i,1:T_i}\) denotes the corresponding reasoning-augmented text target, and \(\theta\) denotes the training model parameters.

We extract the last hidden state corresponding to the \texttt{<eos>} token as the embedding representation. We define \(\mathrm{Emb}(\cdot)\) as the embedding function that maps an input text to its vector representation. Given an input \(x\) and its reasoning text \(r\), the reasoning-augmented embedding is \(\mathrm{Emb}(x \oplus r)\), where \(\oplus\) denotes concatenation. For notational convenience, we define the reasoning-augmented similarity function as follows:
\begin{equation}
\phi(a,b)=\cos(\mathrm{Emb}(a \oplus r_a),\; \mathrm{Emb}(b \oplus r_b))
\label{eq:phi}
\end{equation}
where \(r_a, r_b\) denote the reasoning texts corresponding to \(a\) and \(b\), respectively. Following GTE~\citep{li2023towards}, we adopt the InfoNCE~\citep{oord2018representation} contrastive loss, with the addition of in-batch queries as extra negatives, as defined in Eq.~\eqref{eq:infonce}.
\begin{equation}
\mathcal{L}_{\mathrm{CL}} = -\frac{1}{N} \sum_{i}^{N} \log \frac{e^{\phi(q_i,\, d_{i}^{+})/\tau}}{e^{\phi(q_i,\, d_{i}^{+})/\tau} + \sum_{m=1}^{M} e^{\phi(q_i,\, d_{i,m}^{-})/\tau} + \sum_{j \neq i} e^{\phi(q_i,\, q_{j})/\tau}}
\label{eq:infonce}
\end{equation}
where \(N\) is batch size and \(\tau\) is temperature. \(d_{i}^{+}\) is the positive document, \(d_{i,m}^{-}\) denotes the negative documents, including \(K\) hard negatives and other in-batch documents, and \(q_j\) denotes other in-batch queries.
By jointly optimizing \(\mathcal{L}_{\mathrm{LM}}\) and \(\mathcal{L}_{\mathrm{CL}}\), the model is able to support both text generation and embedding representation within a unified training framework.

\textbf{Reference Model Guidance}:
To preserve the backbone's pretrained embedding competence while restoring generation, we use a frozen copy of the initial embedding backbone as the SFT reference model. For each query \(q\), we collect a candidate document set \(\mathcal{D}\) of one positive and \(K\) hard negatives. 
We align the reasoning-conditioned embedding \(\mathrm{Emb}(x \oplus r)\) with the reference embedding \(\mathrm{Emb}_{\mathrm{ref}}(x)\) produced by this frozen backbone on the original input, using a reference-guided loss \(\mathcal{L}_{\mathrm{RG}}\), as defined in Eq.~\eqref{eq:sft_rg}.
\begin{equation}
\mathcal{L}_{\mathrm{RG}} = \frac{1}{N}\sum_{i=1}^{N} \frac{1}{K+2}\sum_{j=1}^{K+2} \left\|\mathrm{Emb}(x_{i,j} \oplus r_{x_{i,j}})-\mathrm{Emb}_{\mathrm{ref}}(x_{i,j})\right\|_2^2
\label{eq:sft_rg}
\end{equation}
where the inner sum runs over the \(K{+}2\) texts per sample (query plus \(K{+}1\) documents).

\textbf{Overall Reasoning Restoration Objective}:
The overall loss for reasoning restoration stage is:
\begin{equation}
\mathcal{L}_{\mathrm{SFT}}=\mathcal{L}_{\mathrm{LM}}+\mu\mathcal{L}_{\mathrm{CL}}+\gamma\mathcal{L}_{\mathrm{RG}}
\label{eq:sft_all}
\end{equation}
A natural tension exists between $\mathcal{L}_{\mathrm{LM}}$ and $\mathcal{L}_{\mathrm{CL}}$: language modeling allocates capacity toward token-level generation, while contrastive learning demands semantic compression into a single vector. The coefficient $\mu$ controls the contribution of $\mathcal{L}_{\mathrm{CL}}$.
The reference-guided loss $\mathcal{L}_{\mathrm{RG}}$ anchors each reasoning-conditioned embedding to its pre-trained reference to prevent catastrophic forgetting and stabilize representation quality during reasoning restoration, with $\gamma$ controlling its strength.

\subsection{Reasoning Augmentation Reinforcement Learning}
At this stage, we introduce a dual-reward scheme to jointly supervise embedding discrimination and the retrieval utility of generated reasoning (Figure~\ref{fig:framework}, right). An embedding-oriented reward assesses representation quality under reasoning context, preserving the core objective of representation learning. A reasoning-oriented reward encourages generated reasoning to convey retrieval-relevant information that helps distinguish relevant from irrelevant documents, targeting semantic collapse in otherwise fluent generation. We optimize the policy under these rewards using GRPO~\citep{shao2024deepseekmathpushinglimitsmathematical}, which estimates advantages through relative comparisons within sampled output groups without requiring a separate value network.

\textbf{Embedding-Oriented Reward}:
Each batch randomly samples training triples, where documents can also serve as query-side inputs. For each triple \((q, d^{+}, d^{-})\), the current policy samples query reasoning candidates \(\{r_q^i\}\) and generates \(r_{d^{+}}\) and \(r_{d^{-}}\). All reward embeddings are computed by this policy, with no gradient propagation through document-side generation or encoding. For each sampled query reasoning \(r_q^i\), we compute the cosine similarity between the query embedding and both document embeddings, using their margin as the reward:

\begin{equation}
\mathcal{R}_{\mathrm{emb}}(q, r_{q}^{i})=\phi^i(q,\, d^{+}) - \phi^i(q,\, d^{-})
\label{eq:rl_emb}
\end{equation}
where $\phi^{i}$ denotes the similarity function $\phi$ (Eq.~\eqref{eq:phi}) with the query-side reasoning replaced by $r_q^i$.

\textbf{Reasoning-Oriented Reward}:
A discriminative final embedding does not necessarily imply that the generated reasoning is semantically useful for retrieval. We therefore explicitly supervise the alignment between reasoning text and target documents. Using the same inputs as the embedding reward, we employ a pretrained relevance model as the reward model to score query--document pairs augmented with reasoning on both sides. We define the reasoning-oriented reward as the positive--negative score margin: 

\begin{equation}
\mathcal{R}_{\mathrm{reason}}(q,r_q^i)= s(q \oplus r_q^i ,d^{+} \oplus r_{d^{+}}) - s(q \oplus r_q^i ,d^{-} \oplus r_{d^{-}})
\label{eq:rl_rank}
\end{equation}
where \(s(\cdot, \cdot)\) denotes the relevance score produced by the reward model. 
Using a margin rather than a single relevance score prevents a specific form of reward hacking: generic reasoning that raises the positive and negative scores equally cannot increase the reward. The margin instead rewards greater separation between relevant and irrelevant documents. This design also lets the embedding model learn from the reward model's fine-grained relevance signal, improving sensitivity to subtle relevance differences.

\textbf{Embedding Regularization}:
Existing reasoning-augmented embedding models~\citep{gui2025search, lan2025ume} optimize embeddings solely through language-level reasoning during RL, without constraining the embedding output, risking capability degradation. While GRPO's KL penalty prevents generation drift, no analogous constraint exists in the embedding space. We therefore introduce an MSE loss that anchors the policy's embeddings to the reference model's:

\begin{equation}
\mathcal{L}_{\mathrm{MSE}} = \frac{1}{G} \sum_{i=1}^{G} \left\| \mathrm{Emb}_\theta(q \oplus r_q^i) - \mathrm{Emb}_{\theta_{\mathrm{ref}}}(q \oplus r_q^i) \right\|_2^2
\label{eq:rl_mse}
\end{equation}
where $\theta$ and $\theta_{\mathrm{ref}}$ denote the policy and reference model parameters, respectively. The RL reference model is a frozen copy of the model obtained after SFT and is shared by the KL and MSE penalties.

\textbf{Overall Reasoning Augmentation RL Objective}:
The overall loss for RL stage is defined as:
\begin{equation}
\mathcal{L}_{\mathrm{RL}}= \mathcal{L}_{\mathrm{GRPO}}(\lambda \mathcal{R}_{\mathrm{emb}}, \mathcal{R}_{\mathrm{reason}}) + \eta \mathcal{L}_{\mathrm{MSE}}
\label{eq:rl_all}
\end{equation}
$\mathcal{L}_{\mathrm{GRPO}}$ is the policy optimization loss guided by a dual-reward signal (see Appendix~\ref{app:grpo} for detailed formulations). GRPO jointly optimizes embedding discrimination and reasoning relevance, with $\lambda$ weighting the embedding-oriented reward. $\mathcal{L}_{\mathrm{MSE}}$ regularizes the policy model's embeddings toward those of the reference model, with $\eta$ governing its strength. By optimizing this objective, the model learns to jointly maximize both the semantic relevance between reasoning text and query-document pairs and the embedding capability.

\section{Experiments}
\label{sec:exp}

\subsection{Experimental Setup}

\textbf{Training Data.}
We use the bge-en-icl training collection~\citep{flagembedding-dataset}, which covers a range of datasets such as MS MARCO~\citep{msmarco} and ELI5~\citep{eli5}, and supplement it with additional datasets, including Natural Questions~\citep{nq} and TriviaQA~\citep{triviaqa}. Appendix~\ref{app:data_sources} summarizes the data sources and filtering statistics.
The combined corpus contains approximately 7.0M raw instances. After applying the fine-grained reasoning data construction pipeline described in Appendix~\ref{sec:data_pipeline}, we retain approximately 3.6M high-quality query--document tuples. We refer to the resulting dataset as the \textbf{Reasoning Text Embedding Dataset (RTED)} and use it as the common data source for both the SFT and RL stages. SFT uses the filtered reasoning texts as supervision targets, whereas RL uses the original query--document tuples to sample new reasoning texts under the dual-reward objective. We will publicly release RTED to support future research on reasoning-augmented retrieval.

\textbf{Evaluation Benchmarks.}
We evaluate CoFree on the complete Retrieval split of MTEB (English, v2)~\citep{mteb} and all 12 subtasks of BRIGHT~\citep{bright}. MTEB covers a broad range of English retrieval scenarios, including fact verification, argument retrieval, duplicate-question retrieval, multi-hop question answering, and domain-specific retrieval. BRIGHT focuses on reasoning-intensive retrieval across StackExchange, coding, and theorem-based domains. Following the official evaluation protocols, we report nDCG@10 for every dataset and aggregate the dataset-level scores within each benchmark.

\textbf{Baselines.}
We compare CoFree with three categories of embedding models, grouping the results by parameter scale. For \textit{reasoning-based} embeddings, we include Search-R3-Small~\citep{gui2025search} and UME-R1-2B~\citep{lan2025ume}, covering all publicly available models at parameter scales comparable to CoFree. For \textit{decoder-based} embeddings, we include gte-Qwen2-1.5B-instruct~\citep{li2023towards}, stella\_en\_1.5B\_v5~\citep{stella}, udever-bloom-3b~\citep{zhang2023language}, Giga-Embeddings-instruct~\citep{kolodin-ianina-2025-gigaembeddings}, Octen-Embedding-4B~\citep{octen2025rteb}, F2LLM-4B~\citep{F2LLM}, and Qwen3-Embedding-4B~\citep{zhang2025qwen3}. For \textit{encoder-based} embeddings, we include the XL and XXL variants of Sentence-T5~\citep{sentence-t5} and GTR-T5~\citep{gtr-t5}. To control for the effect of training data, we include contrastive fine-tuning baselines (Contrastive FT) at both scales, trained on the same query--document pairs as CoFree using only a contrastive objective, without reasoning augmentation.

\textbf{Implementation Details.}
We train two CoFree variants: CoFree-4B is initialized from Qwen3-Embedding-4B~\citep{zhang2025qwen3}, and CoFree-1.5B is initialized from gte-Qwen2-1.5B-instruct~\citep{li2023towards}. This setup allows us to evaluate the framework across different backbone families and parameter scales. We instantiate the relevance reward model with Qwen3-Reranker-4B~\citep{zhang2025qwen3} to compute the reasoning-oriented reward during RL. \emph{All reasoning-quality evaluators used in our analysis (the three rerankers in Section~\ref{sec:reasoning_quality} and the LLM judges) are distinct from this reward model, avoiding circular evaluation.} To improve SFT efficiency, we group samples by dataset and sequence length, and dynamically adjust the maximum sequence length and per-device batch size for each group (Appendix~\ref{app:batch_grouping}). Complete training hyperparameters are provided in Appendix~\ref{app:impl_details}. Each model uses its own default retrieval instruction, fixed across all evaluation datasets without dataset-specific customization. Experiments are conducted on 32 NVIDIA GPUs, with approximately 200 GB of system memory and 8 CPU cores per node. For CoFree-4B, the SFT and RL stages take approximately 24 and 48 hours, respectively.

\subsection{Overall Performance}
\label{sec:overall_performance}

\begin{table*}[t]
\centering
\scriptsize
\setlength{\tabcolsep}{4pt}
\renewcommand{\arraystretch}{1.0}
\caption{Grouped nDCG@10 results on MTEB English v2 retrieval and BRIGHT. Benchmark Overall columns average their respective datasets; the rightmost Overall averages all 22 datasets (10 MTEB and 12 BRIGHT).}
\vspace{-6pt}
\label{tab:main}
\begin{tabular}{l*{11}{w{c}{0.048\textwidth}}}
\toprule
\multirow{2}{*}{\textbf{Model}} & \multicolumn{6}{c}{\textbf{MTEB (eng, v2)-Retrieval}} & \multicolumn{4}{c}{\textbf{BRIGHT}} & \multirow{2}{*}{\textbf{Overall}} \\
\cmidrule(lr){2-7} \cmidrule(lr){8-11}
& \textbf{Fact} & \textbf{Argu} & \textbf{CQA} & \textbf{MHop} & \textbf{Domain} & \textbf{Overall} & \textbf{Stack} & \textbf{Code} & \textbf{THM} & \textbf{Overall} & \\
\midrule
\multicolumn{12}{c}{\textit{$\sim$1.5B Model Size}} \\
\midrule
Sentence-T5-XL & 26.3 & 42.6 & 50.9 & 41.1 & 38.5 & 39.6 & 20.6 & 14.0 & 6.7 & 16.0 & 26.7 \\
Search-R3-Small\textsuperscript{R} & 34.3 & 48.5 & 44.7 & 41.1 & 42.6 & 42.4 & 15.7 & 7.6 & 6.0 & 11.9 & 25.8 \\
UME-R1-2B\textsuperscript{R} & 39.3 & 38.6 & 39.1 & 35.5 & 30.3 & 36.1 & 11.9 & 8.7 & 8.1 & 10.4 & 22.1 \\
\rowcolor{baselinegray}
gte-Qwen2-1.5B-instruct & 16.6 & 56.2 & 50.5 & 63.8 & 40.3 & 43.1 & 16.4 & 9.4 & 8.8 & 13.4 & 26.9 \\
GTR-T5-XL & 50.6 & 58.3 & 46.2 & 60.7 & 40.0 & 49.1 & 17.4 & 12.3 & 6.4 & 13.8 & 29.8 \\
stella\_en\_1.5B\_v5 & 53.5 & 47.0 & 41.3 & 70.2 & 56.2 & 52.2 & 16.3 & 10.4 & 11.1 & 14.0 & 31.4 \\
Contrastive FT-1.5B & 42.4 & 49.9 & 40.6 & 59.4 & 39.9 & 44.5 & 18.5 & 7.8 & 6.5 & 13.7 & 27.7 \\
\rowcolor{palettegreen!35}
\textbf{(Ours)} CoFree-1.5B & 60.8 & 62.3 & 55.8 & 70.7 & 52.2 & 58.5 & 17.5 & 10.3 & 11.1 & 14.7 & 34.6 \\
\midrule
\multicolumn{12}{c}{\textit{$\sim$4B Model Size}} \\
\midrule
udever-bloom-3b & 14.9 & 31.1 & 46.1 & 12.1 & 45.6 & 33.3 & 6.0 & 14.9 & 1.9 & 6.5 & 18.7 \\
Sentence-T5-XXL & 35.6 & 42.6 & 55.0 & 45.7 & 41.1 & 43.5 & 21.5 & 16.8 & 8.3 & 17.4 & 29.3 \\
GTR-T5-XXL & 51.0 & 58.1 & 47.9 & 61.3 & 38.2 & 49.0 & 18.3 & 11.2 & 5.8 & 14.0 & 29.9 \\
Giga-Embeddings-instruct & 56.3 & 52.8 & 61.7 & 71.4 & 61.1 & 59.6 & 19.9 & 12.9 & 10.3 & 16.3 & 36.0 \\
Octen-Embedding-4B & 58.4 & 68.4 & 61.5 & 68.1 & 55.7 & 61.2 & 20.5 & 9.8 & 17.1 & 17.8 & 37.5 \\
F2LLM-4B & 67.6 & 58.9 & 62.2 & 73.1 & 48.6 & 59.6 & 23.7 & 10.5 & 12.9 & 18.8 & 37.4 \\
\rowcolor{baselinegray}
Qwen3-Embedding-4B & 64.9 & 71.7 & 60.2 & 72.9 & 58.0 & 64.1 & 18.6 & 11.5 & 14.0 & 16.3 & 38.0 \\
Contrastive FT-4B & 48.3 & 51.1 & 52.0 & 61.4 & 48.4 & 50.9 & 21.6 & 8.9 & 12.9 & 17.3 & 32.6 \\
\rowcolor{palettegreen!35}
\textbf{(Ours)} CoFree-4B & 69.5 & 72.8 & 63.4 & 75.8 & 58.8 & 66.4 & 22.2 & 9.9 & 19.4 & 19.4 & 40.8 \\
\bottomrule
\end{tabular}
\par\vspace{3pt}
\begin{minipage}{\textwidth}
\footnotesize\raggedright
Fact = fact verification; Argu = argument retrieval; CQA = CQA duplicate retrieval; MHop = multi-hop QA; Domain = domain-specific retrieval; Stack = StackExchange; Code = coding; THM = theorem-based retrieval. \textsuperscript{R}~denotes reasoning baselines. Gray rows mark the initialization backbones; green rows mark CoFree.
\end{minipage}
\vspace{-10pt}
\end{table*}

\textbf{Main Results.}
Table~\ref{tab:main} summarizes the results on MTEB and BRIGHT. CoFree achieves the best overall performance in both model-size groups: CoFree-4B scores \textbf{40.8}, outperforming Qwen3-Embedding-4B (38.0) by 2.8 points, while CoFree-1.5B scores \textbf{34.6}, surpassing the strongest external baseline, stella\_en\_1.5B\_v5 (31.4), by 3.2 points. CoFree-4B also improves over its backbone on both MTEB (66.4 vs. 64.1) and BRIGHT (19.4 vs. 16.3). CoFree-4B improves on the StackExchange and theorem subsets of BRIGHT. The main exception is BRIGHT-Code, where CoFree-4B scores 9.9 compared with 11.5 for its backbone, whereas CoFree-1.5B improves from 9.4 to 10.3. These results indicate that the effect on code retrieval varies across the evaluated backbones. Appendix~\ref{app:code_retrieval_discussion} discusses the per-task results and possible explanations.

Among reasoning-based models of comparable scale, CoFree-1.5B outperforms Search-R3-Small (25.8) and UME-R1-2B (22.1) by 8.8 and 12.5 overall points, respectively.

Using the same query--document training data, CoFree outperforms its contrastive-only counterpart at both scales: 34.6 vs. 27.7 for 1.5B and 40.8 vs. 32.6 for 4B. At 4B, CoFree-SFT alone already reaches 39.9, compared with 32.6 for contrastive-only fine-tuning (Tables~\ref{tab:main} and~\ref{tab:ablation_components}). These controlled comparisons support the contribution of our training framework beyond simply collecting additional query--document pairs and applying conventional contrastive fine-tuning. Moreover, removing reference guidance from CoFree-SFT reduces the score from 39.9 to 37.1, supporting the contribution of this component within the SFT framework.

\textbf{Inference Efficiency.}
Compared with other reasoning-based embedding models, CoFree-1.5B achieves 4.4--5.4$\times$ the end-to-end throughput of Search-R3-Small and 7.3--12.1$\times$ that of UME-R1-2B, including reasoning generation and embedding extraction. It averages 39.4 and 65.8 generated tokens on MTEB and BRIGHT, respectively.

\textbf{Online Production Verification.} 
We trained CoFree on our industrial dataset and evaluated it through an online A/B test in our e-commerce search system, where it served as an additional recall channel alongside other embedding methods based on Qwen3. 
It achieved absolute gains of 0.02 points in Click-Through Rate (CTR) and 0.12 points in  Conversion Rate (CVR), with relative increases of 13.56\% in orders and 8.15\% in Gross Merchandise Volume (GMV). Figure~\ref{fig:product_results} (Appendix~\ref{appen:online_res_categories}) shows consistent improvements across product categories. CoFree has been fully deployed since January 2026, serving hundreds of millions of users.

\subsection{Reasoning Quality and Collapse}
\label{sec:reasoning_quality}

We examine generation and semantic collapse through complementary evaluations of generation integrity, retrieval utility under frozen encoders, relevance discrimination, and reasoning content. Here, reasoning quality concerns whether generated text preserves input semantics and contributes information useful for relevance discrimination.

\textbf{Mitigating Generation Collapse.}
We assess query- and document-side generation in CoFree-4B before and after SFT using three LLM judges and seq-rep-4~\citep{welleck2019neural}. Table~\ref{tab:generation_collapse} shows that degeneration rates decrease from approximately 100\% to below 4\% across all three judges, while seq-rep-4 drops from 99.37\% to 3.32\%. These results show that SFT substantially reduces degenerate and repetitive generation, mitigating generation collapse. Appendix~\ref{app:generation_collapse} provides the evaluation details.

\suppressfloats[t]
\begin{table}[t]
\centering
\small
\setlength{\tabcolsep}{3pt}
\renewcommand{\arraystretch}{1.05}
\captionsetup{justification=raggedright,singlelinecheck=false}
\begin{minipage}[t]{0.40\textwidth}
\vspace{0pt}
\caption{CoFree-4B generation collapse (\%; lower is better).}
\label{tab:generation_collapse}
\begin{tabular*}{\linewidth}{@{\extracolsep{\fill}}lcc@{}}
\toprule
\textbf{Metric} & \textbf{Before SFT} & \textbf{After SFT} \\
\midrule
Qwen & 99.94 & \textbf{2.81} \\
GLM & 99.96 & \textbf{2.82} \\
Hunyuan & 99.91 & \textbf{3.88} \\
\midrule
seq-rep-4 & 99.37 & \textbf{3.32} \\
\bottomrule
\end{tabular*}
\end{minipage}\hfill
\begin{minipage}[t]{0.56\textwidth}
\vspace{0pt}
\caption{LLM-judge reasoning content scores (1--5; higher is better).}
\label{tab:llm_judge}
\begin{tabular*}{\linewidth}{@{\extracolsep{\fill}}lccc@{}}
\toprule
\textbf{Model} & \textbf{Qwen} & \textbf{GLM} & \textbf{Hunyuan} \\
\midrule
UME-R1 & 2.3331 & 3.2540 & 3.6314 \\
Search-R3 & 2.5645 & 4.0544 & 3.7341 \\
CoFree-1.5B & \textbf{3.3383} & \textbf{4.1014} & \textbf{4.0753} \\
\midrule
\quad w/o reasoning reward & 3.0792 & 3.8746 & 3.8868 \\
\bottomrule
\end{tabular*}
\end{minipage}
\end{table}

\Needspace{6\baselineskip}
\textbf{Transferable Retrieval Utility of Reasoning.}
We assess whether generated reasoning improves retrieval when the encoder is held fixed. Across six encoders, we compare original inputs, Echo and length-matched Noise controls, and reasoning generated by CoFree, Search-R3, and UME-R1. The same generated reasoning is reused across encoders; Appendix~\ref{app:reasoning_controls} provides detailed settings and results.

\begin{figure*}[!htbp]
\centering
\includegraphics[width=\linewidth]{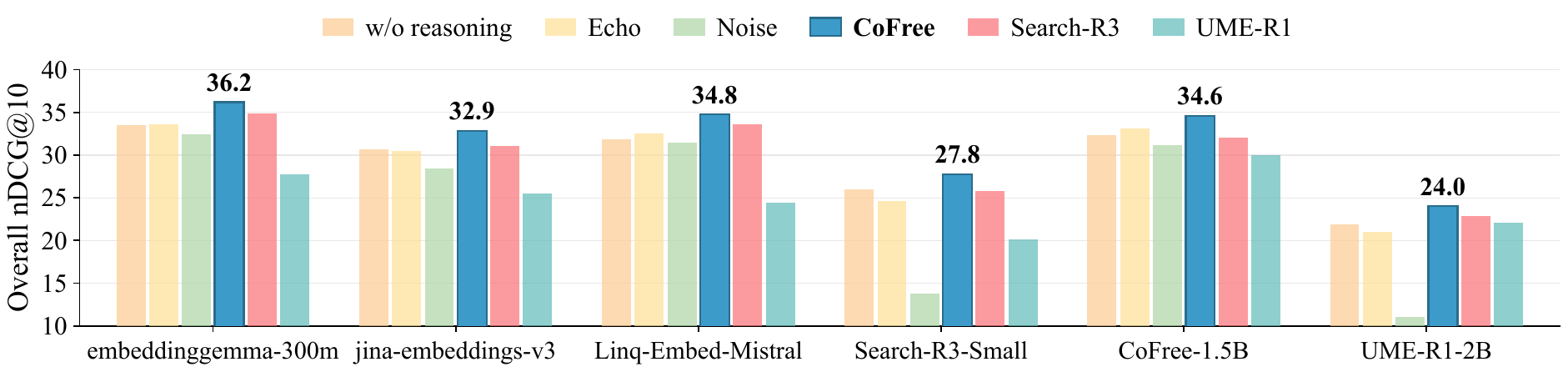}
\caption{Retrieval utility of generated reasoning across frozen encoders. Bars show Overall nDCG@10 for original inputs, Echo/Noise controls, and different reasoning sources.}
\label{fig:motivation}
\vspace{-12pt}
\end{figure*}

Figure~\ref{fig:motivation} shows that UME-R1 reasoning reduces Overall nDCG@10 on five encoders and yields only a small gain on its own encoder. Thus, generated reasoning can impair retrieval utility. In contrast, CoFree reasoning improves Overall nDCG@10 by 1.75--2.94 points over original inputs across all six encoders and outperforms both Echo and Noise controls. These comparisons support the contribution of retrieval-relevant information beyond repetition or uninformative input expansion.

CoFree reasoning also outperforms Search-R3 and UME-R1 reasoning on both MTEB and BRIGHT when evaluated with each baseline's own encoder (Table~\ref{tab:reasoning_controls}). Together, these results support the transferable retrieval utility of CoFree reasoning across both general-purpose and reasoning-based encoders.

\Needspace{14\baselineskip}
\begingroup
\setlength{\columnsep}{12pt}
\setlength{\intextsep}{5pt}
\begin{wrapfigure}{r}{0.46\textwidth}
\centering
\includegraphics[width=\linewidth]{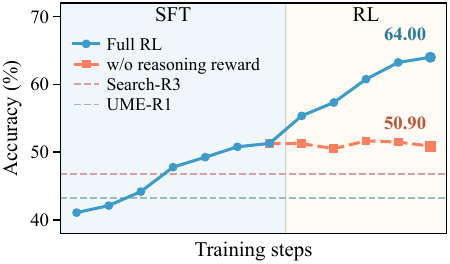}
\captionsetup{justification=justified,singlelinecheck=false,skip=3pt}
\caption{Mean reranker accuracy (\%) across SFT and RL for CoFree-1.5B.}
\label{fig:reranker_checkpoints}
\end{wrapfigure}
\textbf{Improving Reasoning Quality.}
We examine the effect of reasoning supervision on relevance discrimination and reasoning content. Three rerankers distinct from the training reward model assess discrimination. For CoFree-1.5B, their mean pairwise accuracy rises from 51.30\% after SFT to 64.00\% after Full RL, whereas RL without the reasoning reward ends at 50.90\% (Figure~\ref{fig:reranker_checkpoints}). The full model also surpasses Search-R3 (46.83\%) and UME-R1 (43.23\%). These results support the role of explicit reasoning supervision in improving relevance discrimination; Appendix~\ref{app:reranker_diagnostic} provides evaluation details.

On the same aligned query set, Qwen3.5-27B, GLM-4-32B, and Hunyuan-A13B assess intent preservation, reasonable expansion, and specific information contribution on a 1--5 scale, without access to documents. CoFree-1.5B receives higher mean scores than Search-R3 and UME-R1 from all three judges, and exceeds the variant without the reasoning reward by 0.2591, 0.2268, and 0.1885 points, respectively (Table~\ref{tab:llm_judge}; evaluation protocol in Appendix~\ref{app:llm_judge}). Together, the discrimination and content assessments support improved retrieval-relevant semantics, consistent with mitigating semantic collapse.

\par
\endgroup

\Needspace{18\baselineskip}
\subsection{Ablation Study}

\begingroup
\setlength{\columnsep}{12pt}
\setlength{\intextsep}{0pt}
\begin{wraptable}{r}{0.38\textwidth}
\vspace{-3pt}
\centering
\footnotesize
\setlength{\tabcolsep}{4pt}
\renewcommand{\arraystretch}{1.05}
\captionsetup{justification=raggedright,singlelinecheck=false,position=top,aboveskip=0pt,belowskip=3pt}
\caption{CoFree-4B component ablations. $\Delta$: change from the full configuration within each stage.}
\label{tab:ablation_components}
\begin{tabular*}{\linewidth}{@{\extracolsep{\fill}}lcc@{}}
\toprule
\textbf{Configuration} & \textbf{Overall} & $\Delta$ \\
\midrule
\textbf{Full SFT} & 39.9 & - \\
\quad w/o $\mathcal{L}_{\mathrm{RG}}$ & 37.1 & $-2.8$ \\
\quad w/o $\mathcal{L}_{\mathrm{CL}}$ & 38.8 & $-1.1$ \\
\midrule
\textbf{Full RL} & 40.8 & - \\
\quad w/o $\mathcal{R}_{\mathrm{emb}}$ & 40.4 & $-0.4$ \\
\quad w/o $\mathcal{R}_{\mathrm{reason}}$ & 40.3 & $-0.5$ \\
\quad w/o $\mathcal{L}_{\mathrm{MSE}}$ & 40.1 & $-0.7$ \\
\bottomrule
\end{tabular*}
\end{wraptable}

We retain the language modeling loss $\mathcal{L}_{\mathrm{LM}}$ in all SFT configurations to train reasoning generation and ablate the remaining training components.
In the SFT stage (Table~\ref{tab:ablation_components}), the full configuration achieves an overall score of 39.9. Removing reference-guided loss $\mathcal{L}_{\mathrm{RG}}$ causes the largest SFT-stage drop to 37.1 ($-2.8$), supporting its role in preserving representation quality during reasoning activation. Removing $\mathcal{L}_{\mathrm{CL}}$ reduces the score to 38.8 ($-1.1$), showing that direct contrastive supervision provides additional benefits alongside language modeling and reference-guided optimization.

For CoFree-4B, RL improves the full SFT score from 39.9 to 40.8 (+0.9). Removing $\mathcal{R}_{\mathrm{emb}}$ or $\mathcal{R}_{\mathrm{reason}}$ reduces the score by 0.4 and 0.5 points, respectively, supporting the contribution of both rewards to retrieval performance. Removing $\mathcal{L}_{\mathrm{MSE}}$ produces the largest RL-stage drop ($-0.7$), suggesting that explicit embedding-space regularization provides additional benefits beyond the generation-level KL penalty.
Separately, CoFree-1.5B's mean reranker accuracy rises from 51.30\% after SFT to 64.00\% after RL, versus 50.90\% without the reasoning reward (Figure~\ref{fig:reranker_checkpoints}), supporting improved reasoning quality.
\par
\endgroup

\subsection{Training Dynamics and Embedding Structure}
\label{sec:training_dynamics_embeddings}

\begin{figure}[!htbp]
\centering
\includegraphics[width=\linewidth]{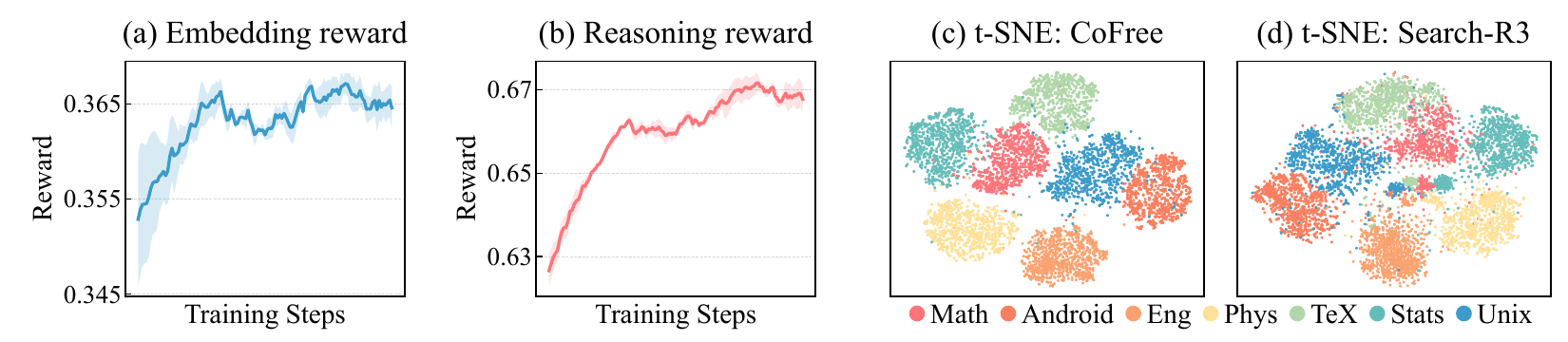}
\captionsetup{skip=4pt}
\caption{RL rewards (mean $\pm$ standard error across three seeds) and t-SNE embeddings on seven CQADupstackRetrieval subdomains.}
\label{fig:reward_tsne}
\vspace{-10pt}
\end{figure}

\textbf{Training Dynamics.}
Both embedding-oriented and reasoning-oriented rewards increase during RL training (Figure~\ref{fig:reward_tsne}(a,b)), indicating progress on both optimization objectives. These trends complement the independent reasoning-quality evaluations in Section~\ref{sec:reasoning_quality}.

\textbf{Embedding Structure.}
CoFree exhibits more distinct domain clusters than Search-R3 in the t-SNE projections (Figure~\ref{fig:reward_tsne}(c,d)), providing a qualitative view of its learned representations. We use seven CQADupstackRetrieval subdomains~\citep{cqa-dup-stack}; sampling and projection settings are provided in Appendix~\ref{app:visualization_settings}. \emph{More experiments and analysis are provided in Appendix~\ref{appen:more_experiments}.}

\section{Conclusion and Future Work}
We present CoFree, a two-stage framework that explicitly optimizes reasoning alongside embeddings to mitigate reasoning collapse. We characterize this collapse through two complementary failure modes: degraded reasoning generation and the loss of retrieval-relevant semantics in otherwise fluent reasoning. Reference-guided SFT aims to restore useful reasoning generation while preserving pretrained embedding competence, and dual-reward RL improves the relevance discrimination of generated reasoning alongside that of the resulting embeddings. This transforms embedding learning from passive encoding into actively reasoned, interpretable retrieval. CoFree-4B achieves an average absolute improvement of 2.8 nDCG@10 points over Qwen3-Embedding-4B across 22 datasets from MTEB and BRIGHT. Experiments on our real-world information retrieval system consistently confirm its gains. We will also release RTED, a 3.6M-instance dataset for reasoning-augmented retrieval. CoFree introduces additional training overhead through its two-stage optimization, teacher-generated demonstrations, and reward-model scoring. Future work will explore streamlined training and more efficient reasoning supervision while preserving retrieval gains.

\subsection*{AI use statement}
We used Qwen3-235B-A22B in non-thinking mode to generate query- and document-side reasoning texts for RTED, following the data construction and automated filtering procedures described in Appendix~\ref{sec:data_pipeline}. We manually spot-checked the generated data. We also used Qwen3-27B to generate synthetic hard negatives for reasoning-quality diagnostics and used language models as automated judges of generation degeneration and reasoning content quality, as detailed in Appendix~\ref{app:full_reasoning_examples}. These model-based assessments are reported separately from retrieval performance measured using standard benchmark corpora and relevance judgments. We also used AI for proofreading. We take responsibility for the final content of this work, including the data, methods, results, and claims.

\bibliographystyle{iclr2027_conference}
\bibliography{references}

\newpage
\appendix

\section{Group Relative Policy Optimization}
\label{app:grpo}

We adopt Group Relative Policy Optimization (GRPO)~\citep{shao2024deepseekmathpushinglimitsmathematical} as the policy optimization algorithm in our RL stage. GRPO replaces the value network in standard PPO~\citep{schulman2017proximal} with group-relative advantage estimation, removing the need for a separate critic and reducing memory overhead.

\textbf{Group Sampling.} For each input prompt $x$, the policy model $\pi_\theta$ samples a group of $G$ candidate responses $\{o_i\}_{i=1}^{G}$. Each response receives a scalar reward $r_i = \lambda\mathcal{R}_{\mathrm{emb}}(x,o_i)+\mathcal{R}_{\mathrm{reason}}(x,o_i)$, combining the rewards defined in Eq.~(\ref{eq:rl_emb}) and Eq.~(\ref{eq:rl_rank}).

\textbf{Group-Relative Advantage.} Instead of estimating value functions, GRPO computes each response's advantage by normalizing its reward against the group statistics:
\begin{equation}
\hat{A}_i = \frac{r_i - \mathrm{mean}(\{r_j\}_{j=1}^{G})}{\mathrm{std}(\{r_j\}_{j=1}^{G})+\epsilon}
\label{eq:grpo_advantage}
\end{equation}
Here, $\epsilon>0$ ensures numerical stability, and the group mean serves as a baseline for favoring above-average responses over below-average ones.

\textbf{Policy Update.} We sample each rollout batch from the current policy and use it for a single parameter update. The batch is not reused. At the start of this update, the importance ratio equals one, so clipping is inactive. The resulting policy-gradient objective includes a KL penalty relative to the frozen SFT reference policy $\pi_{\mathrm{ref}}$ and an MSE penalty in embedding space:
\begin{equation}
\begin{aligned}
\mathcal{L}_{\mathrm{RL}}(\theta)
= -\mathbb{E}_{x,\{o_i\}\sim\pi_\theta}\!\Bigg[
&\frac{1}{\sum_{i=1}^{G}T_i}\sum_{i=1}^{G}\hat{A}_i\log\pi_\theta(o_i\mid x) \\
&-\beta_{\mathrm{KL}}D_{\mathrm{KL}}(\pi_\theta\|\pi_{\mathrm{ref}})
\Bigg]+\eta\mathcal{L}_{\mathrm{MSE}}.
\end{aligned}
\label{eq:grpo_onpolicy_mse}
\end{equation}
The sampled responses and their group-relative advantages remain fixed during the update. For response $o_i$, $T_i$ is the number of valid response tokens, and $\log\pi_\theta(o_i\mid x)$ is the sum of their log probabilities. The policy-gradient term is therefore normalized by the total number of valid response tokens across the group.

The coefficients $\beta_{\mathrm{KL}}$ and $\eta$ control the KL penalty and embedding regularization, respectively. The policy-gradient and KL terms together form $\mathcal{L}_{\mathrm{GRPO}}$ in Eq.~(\ref{eq:rl_all}). The embedding regularizer $\mathcal{L}_{\mathrm{MSE}}$ is defined in Eq.~(\ref{eq:rl_mse}).

\clearpage
\section{Implementation Details}
\label{app:impl_details}

\subsection{Training Configuration}

Table~\ref{tab:impl_details} summarizes the SFT and RL configurations for the main experiments, component ablations, and hyperparameter sensitivity analysis. All settings use AdamW without gradient accumulation, with $K=2$ hard negatives per query.

\emph{SFT.} During SFT, we group samples by dataset and sequence length, dynamically adjusting the maximum sequence length and per-device batch size as described in Appendix~\ref{app:batch_grouping}. At each scale, Contrastive FT uses the same backbone initialization, query--document training pairs, and training settings as the corresponding CoFree-SFT model, with the training objective replaced by the contrastive loss alone.

\emph{RL.} For the CoFree-4B RL component ablations in Table~\ref{tab:ablation_components}, Full RL and all ablated variants start from the same Full SFT checkpoint. Each ablation removes only the specified reward or regularization term, with all other training settings unchanged. The generation-level KL penalty is retained in every RL configuration, including the variant without $\mathcal{L}_{\mathrm{MSE}}$.

\begin{center}
\begin{minipage}{\linewidth}
\captionsetup{type=table}
\centering
\small
\setlength{\tabcolsep}{4pt}
\renewcommand{\arraystretch}{1.2}
\caption{Training hyperparameters for different experimental settings. The Component column lists full-configuration values; each ablation omits the corresponding objective term.}
\label{tab:impl_details}
\begin{tabular}{l *{3}{w{c}{0.12\textwidth}}}
\toprule
\textbf{Hyperparameter} & \textbf{Main} & \textbf{Component} & \textbf{Sensitivity} \\
\midrule
Devices                & 32     & 32             & 16 \\
\midrule
\multicolumn{4}{c}{\textit{SFT Stage}} \\
\midrule
Epochs                                    & 1      & 1              & 3 \\
Learning rate                             & 2e-5   & 2e-5           & 2e-5 \\
Warmup ratio                              & 0.05   & 0.05           & 0.05 \\
$\mu$ ($\mathcal{L}_{\mathrm{CL}}$)       & 1      & 1              & varies \\
$\gamma$ ($\mathcal{L}_{\mathrm{RG}}$)    & 40     & 40             & varies \\
$\tau$ (contrastive temperature)          & 0.07   & 0.07           & 0.07 \\
\midrule
\multicolumn{4}{c}{\textit{RL Stage}} \\
\midrule
Steps                                     & 1000   & 1000           & 300 \\
Learning rate                             & 2e-5   & 2e-5           & 2e-5 \\
Warmup ratio                              & 0.10   & 0.10           & 0.10 \\
$\lambda$ ($\mathcal{R}_{\mathrm{emb}}$)  & 1      & 1              & varies \\
$\beta_{\mathrm{KL}}$ (KL penalty)     & 0.04   & 0.04           & 0.04 \\
$\eta$ ($\mathcal{L}_{\mathrm{MSE}}$)     & 10     & 10             & varies \\
GRPO group size                           & 8      & 8              & 8 \\
Max generation length                     & 1024   & 1024           & 1024 \\
RL sampling temperature                   & 0.99   & 0.99           & 0.99 \\
\bottomrule
\end{tabular}
\end{minipage}
\end{center}

\clearpage
\subsection{Batch Grouping for SFT}
\label{app:batch_grouping}
During SFT, we group samples from the same dataset into batches with similar sequence lengths. Grouping by dataset is intended to make in-batch negatives more informative. Grouping by length allows us to select a maximum sequence length and batch size for each group, improving training efficiency and GPU utilization.

\textbf{Length Estimation.}
We define the length of a training sample as the token count of its longest complete training sequence:
\begin{equation}
l = \max_{x\in\{q,d^+,d^-_1,\ldots,d^-_K\}}\mathrm{len}\!\left(\mathrm{Seq}(x,r_x)\right)
\end{equation}
where $q$ denotes the query, $d^{+}$ the positive document, and $d^{-}_1, \ldots, d^{-}_K$ the hard-negative documents. $\mathrm{Seq}(x,r_x)$ includes the instruction, original text, reasoning target, and all special tokens. We discard SFT samples with $l>3{,}072$.

\textbf{Grouping and Batching.}
We assign each sample to a length bucket using $l$, then form batches from samples that share both a dataset and a bucket. Table~\ref{tab:batch_grouping} lists the maximum sequence length and per-device batch size for each bucket.

\par\smallskip\noindent
\begin{minipage}{\linewidth}
\captionsetup{type=table}
\centering
\small
\setlength{\tabcolsep}{8pt}
\renewcommand{\arraystretch}{1.1}
\caption{Length bucket definitions and corresponding per-device batch sizes for CoFree-1.5B and CoFree-4B.}
\label{tab:batch_grouping}
\begin{tabular}{cccc}
\toprule
\multirow{2}{*}{\textbf{Length Range}} & \multirow{2}{*}{\textbf{Max Seq Len}} & \multicolumn{2}{c}{\textbf{Per-Device Batch Size}} \\
\cmidrule(lr){3-4}
 & & \textbf{CoFree-1.5B} & \textbf{CoFree-4B} \\
\midrule
$(0,\;128]$    & 128  & 72 & 64 \\
$(128,\;256]$  & 256  & 56 & 48 \\
$(256,\;512]$  & 512  & 32 & 24 \\
$(512,\;768]$  & 768  & 16 & 12 \\
$(768,\;1024]$ & 1024 & 12 & 12 \\
$(1024,\;1536]$& 1536 & 8  & 8  \\
$(1536,\;2048]$& 2048 & 6  & 6  \\
$(2048,\;2560]$& 2560 & 4  & 4  \\
$(2560,\;3072]$& 3072 & 4  & 4  \\
\bottomrule
\end{tabular}
\end{minipage}
\par\smallskip

\subsection{Evaluation Protocols}
\label{app:evaluation_protocols}

\emph{Retrieval evaluation.} All results reported in Table~\ref{tab:main} are obtained by re-evaluating each model locally under the same evaluation protocol. We retain each model's default encoding settings, including pooling, and use the default MTEB configurations for embedding normalization and retrieval scoring. For reasoning-augmented inputs, we truncate the concatenated sequence to 4,096 tokens. For CoFree, Search-R3, and UME-R1, we use greedy decoding to generate reasoning chains.

\emph{Inference efficiency.} We measure all compared models on the same NVIDIA GPU hardware using the same inference framework. End-to-end throughput includes both reasoning generation and embedding extraction.

The sampling protocol and results for oracle best-of-$k$ evaluation are provided in Appendix~\ref{app:oracle_best_of_k}.

\begin{samepage}
\subsection{Training Dynamics and Visualization Settings}
\label{app:visualization_settings}
Figure~\ref{fig:reward_tsne}(a,b) reports RL rewards over three random seeds (42, 46, and 123). We smooth each seed's reward trajectory with an exponential moving average (weight 0.95), then plot the mean and standard error across seeds.

For each model in Figure~\ref{fig:reward_tsne}(c,d), we sample 1,000 examples from each of seven CQADupstackRetrieval subdomains~\citep{cqa-dup-stack}, using random seed 42. We fit a separate t-SNE projection for each model with perplexity 30, 1,000 iterations, and random seed 42. Each projection shows the structure within that model's embedding space. Because the projections are fitted independently, their coordinates and distances are not directly comparable across models.

\end{samepage}

\clearpage
\section{More Experiments}
\label{appen:more_experiments}

\subsection{Hyperparameter Sensitivity}
We study how CoFree-4B responds to changes in $\mu$, $\gamma$, $\lambda$, and $\eta$. In each experiment, we vary one coefficient and keep the other settings fixed. We use the reduced training configuration in Appendix~\ref{app:impl_details}, with data sampled from CodeSearchNet, MS MARCO, and ELI5. We evaluate the final checkpoint of each run using the protocol for Table~\ref{tab:main}. Figure~\ref{fig:linear_graph} shows the mean nDCG@10 across 22 datasets, expressed on a 0--1 scale.

\begin{center}
\begin{minipage}{\linewidth}
\captionsetup{type=figure,skip=4pt}
\centerline{\includegraphics[width=\linewidth]{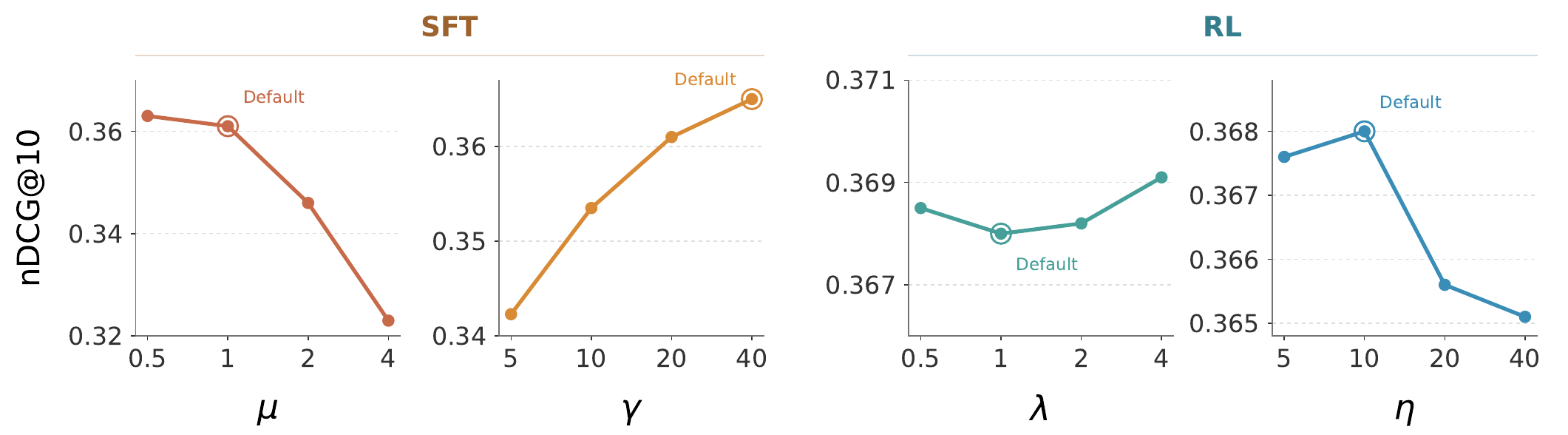}}
\caption{Sensitivity to the SFT-stage loss weights $\mu$ ($\mathcal{L}_{\mathrm{CL}}$) and $\gamma$ ($\mathcal{L}_{\mathrm{RG}}$), and the RL-stage coefficients $\lambda$ ($\mathcal{R}_{\mathrm{emb}}$) and $\eta$ ($\mathcal{L}_{\mathrm{MSE}}$). Performance is measured by nDCG@10. Circled points indicate the default settings.}
\label{fig:linear_graph}
\end{minipage}
\end{center}

\textbf{SFT-stage coefficients.} Performance is similar at $\mu=0.5$ and $\mu=1$, but decreases when $\mu$ is increased to 2 or 4. This suggests that placing too much weight on contrastive learning can hinder joint optimization of the generation and embedding objectives. Increasing $\gamma$ from 5 to 40 consistently improves performance. Under the reduced training configuration used here, stronger reference-guided regularization appears to help preserve embedding quality.

\textbf{RL-stage coefficients.} Performance changes little as $\lambda$ varies over $[0.5,4]$: the maximum and minimum nDCG@10 differ by only 0.0011. This indicates low sensitivity to the embedding-oriented reward weight within the tested range. For $\eta$, performance is similar at 5 and 10, then decreases modestly at 20 and 40. This suggests that overly strong embedding regularization can restrict policy adaptation.

\subsection{Online A/B Results across Categories}
\label{appen:online_res_categories}
In Figure~\ref{fig:product_results}, we show that CoFree achieves consistent improvements across product categories, confirming its broad generalization.

\begin{center}
\begin{minipage}{\linewidth}
\captionsetup{type=figure,skip=4pt}
\centerline{\includegraphics[width=\linewidth]{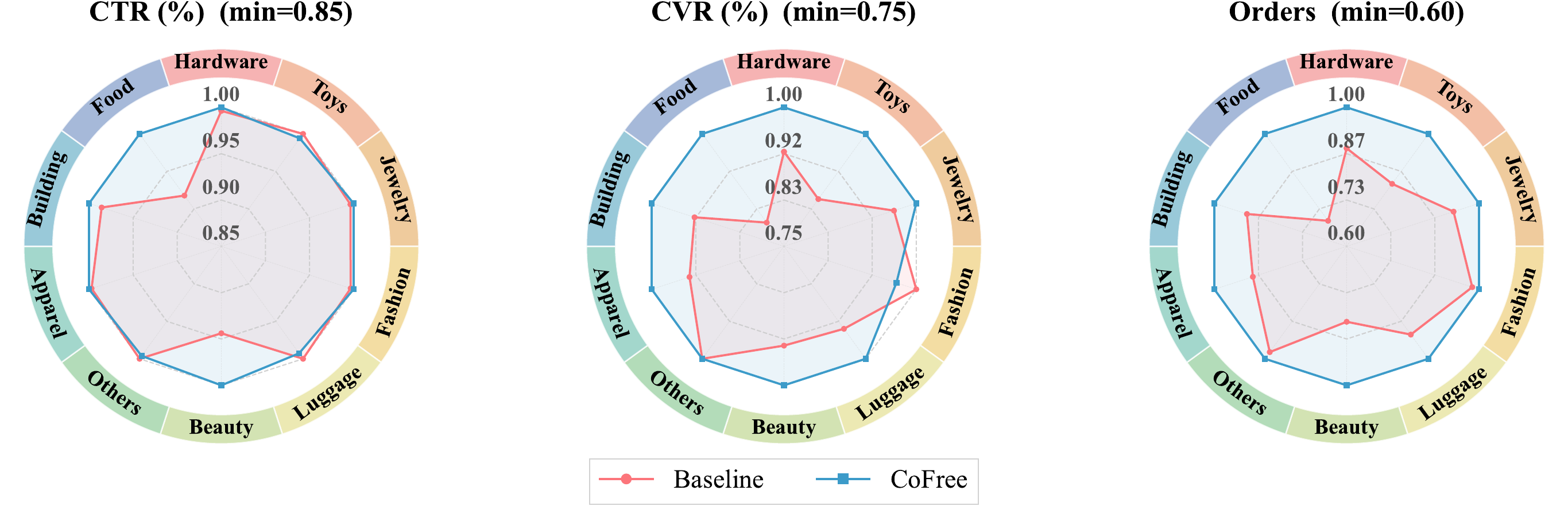}}
\caption{Per-category online A/B test results (CTR, CVR, Orders) across top-10 product categories. Values are normalized per category; inner circle indicates the per-metric minimum ratio.}
\label{fig:product_results}
\end{minipage}
\end{center}

\clearpage
\subsection{Reasoning Transfer across Encoders}
\label{app:reasoning_controls}
We use embeddinggemma-300m~\citep{embedding_gemma_2025}, jina-embeddings-v3~\citep{sturua2024jinaembeddingsv3multilingualembeddingstask}, and Linq-Embed-Mistral~\citep{LinqAIResearch2024} as general-purpose encoders.
The reasoning-based encoders are Search-R3-Small~\citep{gui2025search}, CoFree-1.5B, and UME-R1-2B~\citep{lan2025ume}. Gemma refers to embeddinggemma-300m. The reasoning-source labels CoFree, Search-R3, and UME-R1 denote CoFree-1.5B, Search-R3-Small, and UME-R1-2B, respectively.

Table~\ref{tab:reasoning_controls} compares six input conditions across six fixed encoders. For general-purpose embedding models, we concatenate the original text with the reasoning and retain the official pooling method. Reasoning-based embedding models use their default encoding protocols. All encoders use a maximum sequence length of 4,096 tokens.

We include three control conditions. The first uses only the original query and document text, without adding any reasoning. In \emph{Echo}, we use a copy of the original text in place of the reasoning text. This tests whether repeating information already present in the input can improve retrieval. In \emph{Noise}, we use random tokens in place of reasoning. The tokens are sampled from the vocabulary of the encoder being evaluated. For each input, we match the number of noise tokens to the length of the corresponding CoFree reasoning, measured with that encoder's tokenizer. This tests the effect of adding text of the same length without meaningful reasoning content. Both Echo and Noise are applied to queries and documents.

We compare these controls with reasoning generated by CoFree, Search-R3, and UME-R1. For each source model, we generate the reasoning once and reuse it across all evaluated encoders. Within each encoder, the model remains fixed while the input condition changes. This lets us compare the retrieval performance of different reasoning sources using the same encoder. Evaluating each reasoning-based model with its own reasoning also shows how it performs under its default encoding protocol.

The experiment measures whether generated reasoning improves retrieval and whether its benefit carries over to other encoders. A lower nDCG@10 can arise when the added reasoning is poorly suited to an encoder or causes useful input content to be truncated. A decrease alone therefore does not show that the reasoning has undergone semantic collapse. We assess reasoning quality further through the relevance-discrimination and content evaluations in Appendix~\ref{app:full_reasoning_examples}.

\begin{center}
\begin{minipage}{\linewidth}
\captionsetup{type=table}
\centering
\small
\setlength{\tabcolsep}{4pt}
\renewcommand{\arraystretch}{1.0}
\caption{nDCG@10 (0--100) for reasoning transfer and input controls. Bold: best score per row.}
\label{tab:reasoning_controls}
\begin{tabular*}{\linewidth}{@{\extracolsep{\fill}}lcccccc@{}}
\toprule
\multirow{2}{*}{\textbf{Encoder}} & \multicolumn{3}{c}{\textbf{Input controls}} & \multicolumn{3}{c}{\textbf{Reasoning source}} \\
\cmidrule(lr){2-4}\cmidrule(l){5-7}
 & w/o reasoning & Echo & Noise & \cellcolor{paletteblue!10}\textbf{CoFree} & Search-R3 & UME-R1 \\
\midrule
\rowcolor{black!4}\multicolumn{7}{l}{\textbf{MTEB}} \\
\addlinespace[2pt]
embeddinggemma-300m & 53.63 & 53.74 & 50.38 & \cellcolor{paletteblue!10}\textbf{54.65} & 53.46 & 44.38 \\
jina-embeddings-v3 & 50.23 & 49.52 & 45.08 & \cellcolor{paletteblue!10}\textbf{51.50} & 49.32 & 41.17 \\
Linq-Embed-Mistral & 51.50 & 52.66 & 51.03 & \cellcolor{paletteblue!10}\textbf{53.80} & 52.65 & 40.57 \\
Search-R3-Small & 41.89 & 41.25 & 24.38 & \cellcolor{paletteblue!10}\textbf{43.21} & 42.40 & 33.27 \\
CoFree-1.5B & 55.35 & 56.67 & 53.41 & \cellcolor{paletteblue!10}\textbf{58.50} & 54.22 & 50.78 \\
UME-R1-2B & 35.25 & 34.41 & 18.26 & \cellcolor{paletteblue!10}\textbf{36.73} & 36.15 & 36.10 \\
\addlinespace[4pt]
\rowcolor{black!4}\multicolumn{7}{l}{\textbf{BRIGHT}} \\
\addlinespace[2pt]
embeddinggemma-300m & 16.73 & 16.92 & 17.53 & \cellcolor{paletteblue!10}\textbf{20.83} & 19.41 & 13.95 \\
jina-embeddings-v3 & 14.43 & 14.67 & 14.57 & \cellcolor{paletteblue!10}\textbf{17.31} & 15.88 & 12.52 \\
Linq-Embed-Mistral & 15.45 & 15.68 & 15.19 & \cellcolor{paletteblue!10}\textbf{18.93} & 17.81 & 10.93 \\
Search-R3-Small & 12.80 & 10.84 & 4.89 & \cellcolor{paletteblue!10}\textbf{14.90} & 11.90 & 9.26 \\
CoFree-1.5B & 13.14 & 13.55 & 12.59 & \cellcolor{paletteblue!10}\textbf{14.70} & 13.58 & 12.66 \\
UME-R1-2B & 10.70 & 9.75 & 5.09 & \cellcolor{paletteblue!10}\textbf{13.47} & 11.88 & 10.40 \\
\addlinespace[4pt]
\rowcolor{black!4}\multicolumn{7}{l}{\textbf{Overall (10 MTEB + 12 BRIGHT datasets)}} \\
\addlinespace[2pt]
embeddinggemma-300m & 33.50 & 33.66 & 32.46 & \cellcolor{paletteblue!10}\textbf{36.20} & 34.89 & 27.78 \\
jina-embeddings-v3 & 30.70 & 30.51 & 28.44 & \cellcolor{paletteblue!10}\textbf{32.85} & 31.08 & 25.54 \\
Linq-Embed-Mistral & 31.84 & 32.49 & 31.48 & \cellcolor{paletteblue!10}\textbf{34.78} & 33.65 & 24.40 \\
Search-R3-Small & 26.02 & 24.66 & 13.75 & \cellcolor{paletteblue!10}\textbf{27.77} & 25.76 & 20.17 \\
CoFree-1.5B & 32.33 & 33.15 & 31.14 & \cellcolor{paletteblue!10}\textbf{34.61} & 32.05 & 29.99 \\
UME-R1-2B & 21.86 & 20.96 & 11.08 & \cellcolor{paletteblue!10}\textbf{24.04} & 22.91 & 22.08 \\
\bottomrule
\end{tabular*}
\end{minipage}
\end{center}

\clearpage
\subsection{Oracle Best-of-\texorpdfstring{$k$}{k} Analysis}
\label{app:oracle_best_of_k}

We examine how retrieval performance varies across sampled reasoning trajectories for CoFree-1.5B and CoFree-4B on ArguAna, CQADupstackGaming, CQADupstackUnix, and SCIDOCS. We run 64 independent retrieval passes. In each pass, we separately sample reasoning for queries and documents at temperature 0.9.

For each query and each $k\in\{1,2,4,8,16,32,64\}$, we randomly select $k$ of these passes. We use the ground-truth relevance judgments to score their rankings and retain the highest nDCG@10 for that query. We then average these best scores across queries. To estimate the expected best-of-$k$ score, we repeat this selection and averaging procedure for 1,000 Monte Carlo trials at each $k$. All trials reuse the same pool of 64 retrieval passes; they require no additional model inference or independent training runs. The $k=1$ point estimates performance for a single sample at temperature 0.9, whereas the main retrieval results use greedy decoding.

\begin{center}
\begin{minipage}{\linewidth}
\centering
\captionsetup{type=figure,skip=4pt}
\includegraphics[width=\linewidth]{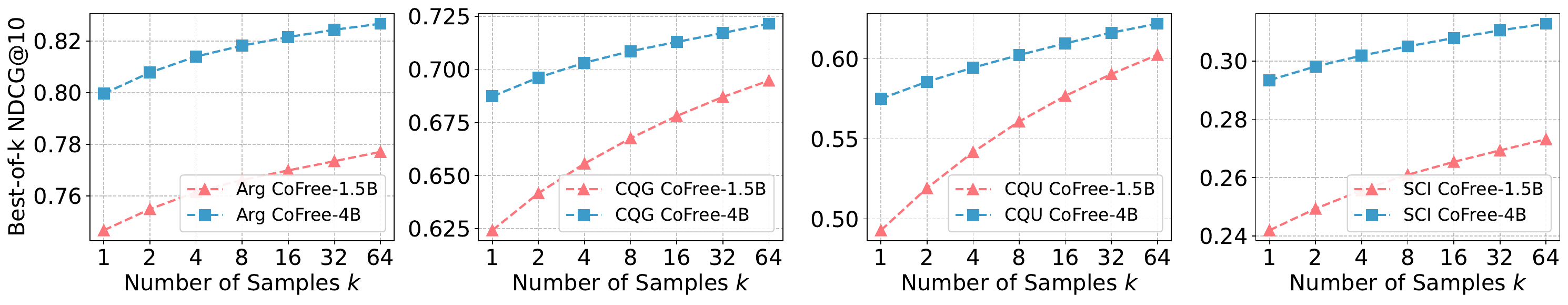}
\caption{Oracle best-of-$k$ nDCG@10 with sampled reasoning for CoFree-1.5B and CoFree-4B.}
\label{fig:passk}
\end{minipage}
\end{center}

Figure~\ref{fig:passk} shows oracle best-of-$k$ gains on all four datasets at both model sizes, with particularly large gains for CoFree-1.5B on the duplicate-question retrieval benchmarks. These results show that different sampled reasoning trajectories can lead to different retrieval outcomes. The gap between single-sample and oracle performance suggests an opportunity to improve retrieval by identifying and selecting more useful reasoning trajectories.

\subsection{Discussion of Code Retrieval}
\label{app:code_retrieval_discussion}

Table~\ref{tab:detailed_bright} shows that CoFree-4B scores below its backbone on LeetCode (21.2$\rightarrow$18.7) and Pony (1.9$\rightarrow$1.1). Contrastive-only fine-tuning on the same query--document pairs, without reasoning augmentation, also reduces the scores on these datasets, to 16.4 and 1.4, respectively. The decline therefore occurs under both training settings, although CoFree has a higher aggregate Code score than the contrastive-only baseline (9.9 vs. 8.9).

For CoFree-1.5B, the aggregate Code score improves from 9.4 to 10.3. This combines a gain on LeetCode (15.9$\rightarrow$18.4) with a decline on Pony (2.9$\rightarrow$2.1). The aggregate outcome thus differs between the two backbones, while performance on Pony declines for both.

One possible explanation is that continued training on the mixed-domain corpus changes the code-retrieval capabilities learned by the backbone. Limited coverage of task-relevant code reasoning in the retained supervision may also contribute. Table~\ref{tab:data_source_filtering} shows that CodeSearchNet retains 186,058 instances, or 13.54\% of its original pool, and CoSQA contributes 13,293 retained instances. Together, these two code-oriented sources account for approximately 5.5\% of the retained tuples. These observations motivate examining both the retention of the backbone's capabilities and the coverage of code-specific supervision. The relative contribution of each factor to the decline remains to be established.

\clearpage
\section{RTED Data Construction}
\label{sec:data_pipeline}

Figure~\ref{fig:data_process} outlines the RTED construction pipeline, from query--document pairs to filtered training tuples with reasoning text.

\begin{center}
\begin{minipage}{\linewidth}
\centering
\captionsetup{type=figure,skip=4pt}
\includegraphics[width=1.0\linewidth]{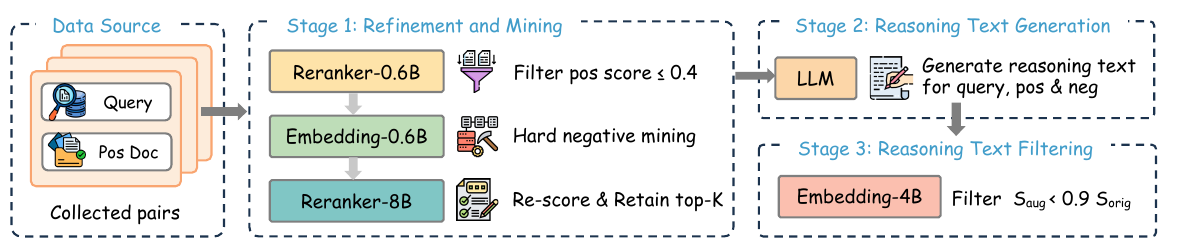}
\caption{Overview of the RTED construction pipeline for SFT and RL.}
\label{fig:data_process}
\end{minipage}
\end{center}

\subsection{Data Sources and Filtering Statistics}
\label{app:data_sources}

Table~\ref{tab:data_source_filtering} summarizes the training data composition and filtering statistics. All datasets use their training splits.

\begin{table}[!htbp]
\centering
\footnotesize
\setlength{\tabcolsep}{4pt}
\renewcommand{\arraystretch}{1.15}
\caption{Training data composition and filtering statistics. Retention is the percentage of instances retained after filtering.}
\label{tab:data_source_filtering}
\begin{tabular*}{\textwidth}{@{\extracolsep{\fill}}l c r r r@{}}
\toprule
\textbf{Dataset} & \textbf{Split} & \makecell{\textbf{Before}\\\textbf{filtering}} & \makecell{\textbf{After}\\\textbf{filtering}} & \makecell{\textbf{Retention}\\\textbf{(\%)}} \\
\midrule
bge-en-icl~\citep{flagembedding-dataset} & train & 2,188,119 & 1,604,900 & 73.35 \\
AG News~\citep{agnews} & train & 167,677 & 62,895 & 37.51 \\
Amazon QA~\citep{amazon-qa} & train & 579,165 & 413,143 & 71.33 \\
CodeSearchNet~\citep{codesearchnet} & train & 1,373,821 & 186,058 & 13.54 \\
CoSQA~\citep{cosqa} & train & 19,604 & 13,293 & 67.81 \\
GooAQ~\citep{gooaq} & train & 416,681 & 201,755 & 48.42 \\
HotpotQA~\citep{hotpotqa-hard-negatives} & train & 169,032 & 152,709 & 90.34 \\
Natural Questions~\citep{nq} & train & 58,568 & 47,104 & 80.43 \\
SciFact~\citep{scifact} & train & 919 & 445 & 48.42 \\
\makecell[l]{StackExchange (Duplicate Posts)\\\citep{stackexchange-duplicates}} & train & 250,517 & 155,155 & 61.93 \\
\makecell[l]{StackExchange (Voted Answers)\\\citep{StackExchangeDataset}} & train & 1,049,481 & 414,465 & 39.49 \\
TriviaQA~\citep{triviaqa} & train & 731,862 & 361,701 & 49.42 \\
\bottomrule
\end{tabular*}
\end{table}

\subsection{Construction Pipeline}
\label{app:rted_construction_pipeline}

The pipeline in Figure~\ref{fig:data_process} has three stages: refinement and mining, reasoning text generation, and reasoning text filtering. It produces approximately 3.6M retained training instances from 7.0M raw instances. Each retained instance is a query--document tuple
\begin{equation}
z_i = \left(q_i, d_i^+, \{d_{i,k}^-\}_{k=1}^{K}\right), \qquad K=2,
\label{eq:rted_tuple}
\end{equation}
containing one query, one positive document, and two hard-negative documents. Each query and document has associated reasoning text.

\textbf{Refinement and Mining.}
We first remove noisy or weakly related positive pairs using Qwen3-Reranker-0.6B~\citep{zhang2025qwen3}. Let $s_{0.6\mathrm B}(q,d)$ and $s_{8\mathrm B}(q,d)$ denote the relevance scores assigned to a query--document pair by Qwen3-Reranker-0.6B and Qwen3-Reranker-8B, respectively. A positive pair is retained only if
\begin{equation}
s_{0.6\mathrm B}(q,d^+) \geq 0.4.
\label{eq:rted_positive_filter}
\end{equation}

\Needspace{6\baselineskip}
For samples that do not already have negative documents, we retrieve candidates from the same sub-dataset using Qwen3-Embedding-0.6B~\citep{zhang2025qwen3}. We exclude all known positive documents for the query. Among the top 30 retrieved documents, we randomly select 10 from ranks 10 through 30, inclusive. These documents form the hard-negative candidate set $\mathcal C(q)$.

We then re-score these candidates and the corresponding positive pair with Qwen3-Reranker-8B. To reduce false negatives, we filter candidates using the positive-pair score as a reference:
\begin{equation}
\mathcal C_{\mathrm{valid}}(q)
= \left\{d\in\mathcal C(q):
s_{8\mathrm B}(q,d) \leq 0.95\,s_{8\mathrm B}(q,d^+)\right\}.
\label{eq:rted_negative_filter}
\end{equation}
We retain the $K=2$ highest-scoring candidates in $\mathcal C_{\mathrm{valid}}(q)$ as hard negatives and discard samples with fewer than $K$ eligible candidates.

\textbf{Fine-Grained Reasoning Text Generation.}
We use Qwen3-235B-A22B~\citep{qwen3technicalreport} in non-thinking mode to generate reasoning text. Each query or document is processed independently, using only its own content. For queries, the teacher clarifies the retrieval intent, completes fragmentary expressions, and enriches the query with relevant contextual terms and synonyms to broaden retrieval coverage. These expansions preserve the original intent and constraints. For documents, it makes factual relationships explicit and removes irrelevant wording. The prompts instruct the teacher to preserve the original meaning, named entities, numerical values, and factual qualifiers. The complete prompts appear in Appendix~\ref{app:reason_prompt}.

For each input text $x$, the teacher generates reasoning text $r_x$, which serves as the SFT supervision target and is concatenated with $x$ for embedding learning.

\begin{samepage}
\textbf{Reasoning Text Filtering.}
We filter the generated reasoning by checking how it changes the similarity between each query and its positive document. Let $E_{\mathrm f}$ denote the fixed Qwen3-Embedding-4B encoder used for this step. We first compute the cosine similarity of the original query and positive document:
\begin{equation}
s_{\mathrm{orig}}
= \cos\!\left(E_{\mathrm f}(q),\,E_{\mathrm f}(d^+)\right).
\label{eq:rted_original_similarity}
\end{equation}
\end{samepage}
We then concatenate each input text with its associated reasoning text and compute the augmented similarity using the same encoder:
\begin{equation}
s_{\mathrm{aug}}
= \cos\!\left(E_{\mathrm f}(q\oplus r_q),\,
E_{\mathrm f}(d^+\oplus r_{d^+})\right),
\label{eq:rted_augmented_similarity}
\end{equation}
where $\oplus$ denotes text concatenation. A tuple is retained only if
\begin{equation}
s_{\mathrm{aug}} \geq 0.9\,s_{\mathrm{orig}}.
\label{eq:rted_similarity_filter}
\end{equation}
This rule compares the similarity after adding reasoning with the similarity of the original pair. We discard tuples that do not meet the threshold.

The retained tuples and their reasoning text form RTED. Both training stages use this dataset, but they use the reasoning differently. SFT learns from the teacher-generated reasoning text. RL uses the query--document tuples to sample new reasoning from the current policy and trains with the dual-reward objective.

\clearpage
\subsection{Instructions for Reasoning Text Generation}
\label{app:reason_prompt}

The following prompts are used to generate the teacher reasoning text for SFT. The query prompt asks the teacher to make the retrieval intent explicit and add relevant contextual terms and synonyms to broaden retrieval coverage. The document prompt asks it to clarify the factual content. Both prompts require preservation of the original meaning and key information.

\begin{paperprompt}{Query Reasoning Text Generation}{prompt:query_side}
You are a semantic augmentation engine for information retrieval queries.

Your task:
Generate reasoning text for the input query to support semantic retrieval against a knowledge base. Make the retrieval intent explicit while preserving the original meaning. Use wording distinct from the input and write in the same language as the input.

Core Principles:
1. **Repair and complete**: Fix typos, expand bare keywords into complete natural language questions. e.g., "battery life" -> "What is the battery life of this device?"
2. **Make intent explicit**: Identify the query intent (e.g., compatibility, specification, procedural, troubleshooting) and ensure the reasoning text unambiguously reflects it. Do not shift between intents.
3. **Normalize vague references**: Resolve ambiguous pronouns (it / this) with the specific subject or "this item" if context is absent. Expand common abbreviations.
4. **Allow minor expansion**: You may add relevant context words or synonyms to broaden retrieval coverage, provided the core intent remains unchanged.
5. **Preserve exactly**: Named entities, technical terms, and specific identifiers must remain character-for-character. Never correct, expand, or normalize them.

\end{paperprompt}

\newpage

\begin{paperprompt}{Document Reasoning Text Generation}{prompt:doc_side}
You are a semantic augmentation engine for information retrieval answer passages.

Your task:
Generate reasoning text for the input answer passage to support semantic retrieval against user questions. Make the key facts and their relationships explicit while preserving the original meaning. For every input, produce reasoning text with wording distinct from the input and write in the same language as the input.

Core Principles:
1. **Anchor bare facts**: If an attribute value is stated without its attribute name, make the relationship explicit. e.g., "6 feet." -> "The length is 6 feet."
2. **Convert first-person to third-person**: Extract factual claims from personal experience framing.
3. **Preserve verdicts and hedges exactly**: Never alter the polarity or certainty of compatibility, safety, or experiential claims. Retain qualifiers ("seems to", "in my experience", "only if") as-is.
4. **Strip noise**: Remove filler ("Hope this helps!", "Great product!"), emoji, informal shorthand (idk -> I do not know, bc -> because), promotional boilerplate, and platform-specific metadata.
5. **Preserve exactly**: All numerical values, units, brand names, model numbers, and technical identifiers character-for-character.

\end{paperprompt}

\Needspace{6\baselineskip}
\section{Detailed Benchmark Results}
\label{app:detailed_results}
Tables~\ref{tab:detailed_mteb} and~\ref{tab:detailed_bright} report per-dataset nDCG@10 scores on the 10 MTEB English v2 retrieval datasets and 12 BRIGHT sub-datasets, respectively.

\begin{table*}[!htbp]
\centering
\scriptsize
\setlength{\tabcolsep}{1.5pt}
\renewcommand{\arraystretch}{1.15}
\caption{Detailed nDCG@10 results on MTEB English v2 retrieval. Task categories follow Table~\ref{tab:main}; higher is better.}
\label{tab:detailed_mteb}
\settowidth{\benchmarkmodelwidth}{Giga-Embeddings-instruct}
\begin{tabular}{@{}l*{11}{w{c}{\dimexpr(\textwidth-\benchmarkmodelwidth-22\tabcolsep)/11\relax}}@{}}
\toprule
\multirow{2}{*}{\textbf{Model}} & \multicolumn{2}{c}{\textbf{Fact}} & \multicolumn{2}{c}{\textbf{Argu}} & \multicolumn{2}{c}{\textbf{CQA}} & \multicolumn{1}{c}{\textbf{MHop}} & \multicolumn{3}{c}{\textbf{Domain}} & \multirow{2}{*}{\textbf{Avg.}} \\
\cmidrule(lr){2-3} \cmidrule(lr){4-5} \cmidrule(lr){6-7} \cmidrule(lr){8-8} \cmidrule(lr){9-11}
& \makecell{Climate\\FEVER\\HN} & \makecell{FEVER\\HN} & \makecell{ArguAna} & \makecell{Touche\\2020\\v3} & \makecell{CQA\\Gaming} & \makecell{CQA\\Unix} & \makecell{Hotpot\\QA\\HN} & \makecell{FiQA\\2018} & \makecell{SCI\\DOCS} & \makecell{TREC\\COVID} & \\
\midrule
\multicolumn{12}{c}{\textit{$\sim$1.5B Model Size}} \\
\midrule
Sentence-T5-XL & 11.6 & 41.0 & 39.4 & 45.7 & 59.9 & 41.8 & 41.1 & 44.7 & 16.0 & 54.8 & 39.6 \\
Search-R3-Small\textsuperscript{R} & 20.8 & 47.7 & 47.5 & 49.6 & 49.8 & 39.7 & 41.1 & 32.5 & 19.1 & 76.1 & 42.4 \\
UME-R1-2B\textsuperscript{R} & 21.7 & 56.9 & 47.1 & 30.0 & 45.9 & 32.3 & 35.5 & 27.3 & 15.8 & 48.0 & 36.1 \\
\rowcolor{baselinegray}
gte-Qwen2-1.5B-instruct & 11.4 & 21.9 & 60.7 & 51.7 & 58.6 & 42.5 & 63.8 & 43.9 & 22.0 & 55.0 & 43.1 \\
GTR-T5-XL & 27.6 & 73.5 & 52.9 & \textbf{63.7} & 55.8 & 36.6 & 60.7 & 44.2 & 15.7 & 60.1 & 49.1 \\
stella\_en\_1.5B\_v5 & 28.4 & 78.6 & 57.0 & 36.9 & 53.6 & 29.0 & 70.2 & \textbf{57.7} & \textbf{26.8} & 84.1 & 52.2 \\
Contrastive FT-1.5B & 24.6 & 60.1 & 67.2 & 32.6 & 44.1 & 37.1 & 59.4 & 32.8 & 16.8 & 70.0 & 44.5 \\
\rowcolor{palettegreen!35}
CoFree-1.5B & \textbf{32.1} & \textbf{89.5} & \textbf{74.6} & 50.0 & \textbf{62.4} & \textbf{49.2} & \textbf{70.7} & 47.7 & 24.3 & \textbf{84.5} & \textbf{58.5} \\
\midrule
\multicolumn{12}{c}{\textit{$\sim$4B Model Size}} \\
\midrule
udever-bloom-3b & 9.0 & 20.8 & 48.3 & 13.9 & 54.7 & 37.4 & 12.1 & 35.1 & 18.5 & 83.2 & 33.3 \\
Sentence-T5-XXL & 15.2 & 55.9 & 39.9 & 45.3 & 63.5 & 46.6 & 45.7 & 46.7 & 17.2 & 59.5 & 43.5 \\
GTR-T5-XXL & 27.3 & 74.7 & 53.8 & 62.4 & 57.5 & 38.4 & 61.3 & 46.8 & 15.9 & 51.9 & 49.0 \\
Giga-Embeddings-instruct & 28.0 & 84.6 & 47.2 & 58.4 & \textbf{70.2} & 53.3 & 71.4 & \textbf{82.4} & 19.9 & 81.0 & 59.6 \\
Octen-Embedding-4B & 33.6 & 83.2 & 71.9 & 64.9 & 68.2 & 54.7 & 68.1 & 59.1 & 25.7 & 82.3 & 61.2 \\
F2LLM-4B & 43.4 & 91.8 & 61.9 & 55.9 & 65.4 & \textbf{59.0} & 73.1 & 58.4 & 26.7 & 60.6 & 59.6 \\
\rowcolor{baselinegray}
Qwen3-Embedding-4B & 42.5 & 87.3 & 69.6 & \textbf{73.8} & 66.5 & 53.8 & 72.9 & 58.8 & 27.0 & 88.4 & 64.1 \\
Contrastive FT-4B & 24.7 & 71.9 & 67.4 & 34.8 & 54.5 & 49.5 & 61.4 & 44.6 & 24.0 & 76.6 & 50.9 \\
\rowcolor{palettegreen!35}
CoFree-4B & \textbf{46.6} & \textbf{92.5} & \textbf{80.3} & 65.2 & 69.0 & 57.8 & \textbf{75.8} & 58.3 & \textbf{29.5} & \textbf{88.7} & \textbf{66.4} \\
\bottomrule
\end{tabular}
\vspace{3pt}
\begin{minipage}{\textwidth}
\footnotesize\raggedright
Bold marks the best score within each size group, including ties; gray and green rows mark initialization backbones and CoFree, respectively. Avg. is the dataset-level mean. \textsuperscript{R}~denotes reasoning baselines. Full MTEB dataset identifiers (in column order): ClimateFEVERHardNegatives, FEVERHardNegatives, ArguAna, Touche2020Retrieval.v3, CQADupstackGamingRetrieval, CQADupstackUnixRetrieval, HotpotQAHardNegatives, FiQA2018, SCIDOCS, and TRECCOVID. HN denotes HardNegatives; v3 denotes the retrieval dataset version.
\end{minipage}
\end{table*}

\begin{table*}[!htbp]
\centering
\scriptsize
\setlength{\tabcolsep}{1.5pt}
\renewcommand{\arraystretch}{1.15}
\caption{Detailed nDCG@10 results on BRIGHT. Task categories follow Table~\ref{tab:main}; higher is better.}
\label{tab:detailed_bright}
\settowidth{\benchmarkmodelwidth}{Giga-Embeddings-instruct}
\begin{tabular}{@{}l*{13}{w{c}{\dimexpr(\textwidth-\benchmarkmodelwidth-26\tabcolsep)/13\relax}}@{}}
\toprule
\multirow{2}{*}{\textbf{Model}} & \multicolumn{7}{c}{\textbf{Stack}} & \multicolumn{2}{c}{\textbf{Code}} & \multicolumn{3}{c}{\textbf{THM}} & \multirow{2}{*}{\textbf{Avg.}} \\
\cmidrule(lr){2-8} \cmidrule(lr){9-10} \cmidrule(lr){11-13}
& \makecell{Bio.} & \makecell{Earth\\Sci.} & \makecell{Econ.} & \makecell{Psych.} & \makecell{Robot.} & \makecell{Stack\\Over.} & \makecell{Sust.\\Living} & \makecell{Leet\\Code} & \makecell{Pony} & \makecell{AoPS} & \makecell{TQA\\Ques.} & \makecell{TQA\\Thm.} & \\
\midrule
\multicolumn{14}{c}{\textit{$\sim$1.5B Model Size}} \\
\midrule
Sentence-T5-XL & \textbf{22.0} & 24.0 & \textbf{23.0} & \textbf{28.4} & 14.8 & 11.1 & \textbf{20.6} & 14.8 & \textbf{13.2} & 1.5 & 10.2 & 8.3 & \textbf{16.0} \\
Search-R3-Small\textsuperscript{R} & 15.8 & 21.6 & 16.3 & 20.0 & 10.7 & 13.8 & 11.7 & 13.3 & 1.8 & 0.1 & 8.0 & 9.8 & 11.9 \\
UME-R1-2B\textsuperscript{R} & 11.5 & 18.8 & 11.6 & 12.3 & 10.3 & 9.5 & 9.6 & 16.3 & 1.2 & 2.5 & 14.0 & 7.7 & 10.4 \\
\rowcolor{baselinegray}
gte-Qwen2-1.5B-instruct & 14.0 & 27.5 & 19.2 & 21.6 & 7.7 & 7.3 & 17.9 & 15.9 & 2.9 & \textbf{5.9} & 12.5 & 8.1 & 13.4 \\
GTR-T5-XL & 15.4 & 27.2 & 15.3 & 20.1 & \textbf{15.1} & 12.3 & 16.2 & 16.9 & 7.8 & 2.8 & 9.3 & 7.1 & 13.8 \\
stella\_en\_1.5B\_v5 & 18.6 & 27.1 & 15.2 & 15.9 & 13.2 & 9.1 & 15.2 & \textbf{19.1} & 1.6 & 4.8 & 13.5 & \textbf{14.9} & 14.0 \\
Contrastive FT-1.5B & 19.9 & 26.2 & 18.1 & 19.2 & 12.9 & \textbf{13.9} & 19.3 & 15.1 & 0.5 & 2.2 & 13.7 & 3.7 & 13.7 \\
\rowcolor{palettegreen!35}
CoFree-1.5B & 17.1 & \textbf{27.9} & 17.1 & 20.7 & 10.8 & 11.8 & 17.1 & 18.4 & 2.1 & 5.0 & \textbf{16.4} & 11.9 & 14.7 \\
\midrule
\multicolumn{14}{c}{\textit{$\sim$4B Model Size}} \\
\midrule
udever-bloom-3b & 7.1 & 5.9 & 7.3 & 4.7 & 3.8 & 6.3 & 6.8 & 15.8 & 13.9 & \textbf{3.9} & 1.9 & 0.0 & 6.5 \\
Sentence-T5-XXL & 26.7 & 25.9 & \textbf{24.2} & 28.4 & 12.8 & 10.8 & \textbf{21.5} & 16.8 & \textbf{16.7} & 2.2 & 10.9 & 11.7 & 17.4 \\
GTR-T5-XXL & 15.7 & 26.0 & 17.5 & 21.3 & \textbf{17.7} & 14.1 & 15.6 & 16.6 & 5.7 & 2.4 & 8.8 & 6.3 & 14.0 \\
Giga-Embeddings-instruct & 22.9 & 33.6 & 17.1 & 22.0 & 14.0 & 14.8 & 15.1 & 17.8 & 8.0 & 2.3 & 16.0 & 12.5 & 16.3 \\
Octen-Embedding-4B & 21.8 & 31.5 & 17.8 & 24.5 & 13.7 & 16.3 & 17.6 & 17.4 & 2.1 & 2.6 & 18.8 & 29.8 & 17.8 \\
F2LLM-4B & 24.5 & \textbf{35.7} & 23.9 & \textbf{29.8} & 16.2 & \textbf{17.7} & 18.3 & 19.3 & 1.7 & 2.7 & 18.1 & 18.0 & 18.8 \\
\rowcolor{baselinegray}
Qwen3-Embedding-4B & 16.1 & 31.1 & 15.4 & 22.2 & 14.9 & 15.0 & 15.9 & \textbf{21.2} & 1.9 & 3.2 & 16.4 & 22.5 & 16.3 \\
Contrastive FT-4B & \textbf{28.7} & 30.5 & 21.8 & 28.9 & 12.7 & 9.5 & 19.2 & 16.4 & 1.4 & 3.8 & 21.4 & 13.5 & 17.3 \\
\rowcolor{palettegreen!35}
CoFree-4B & 22.4 & 33.6 & 21.6 & 26.3 & 16.0 & 16.9 & 18.5 & 18.7 & 1.1 & 2.9 & \textbf{22.6} & \textbf{32.6} & \textbf{19.4} \\
\bottomrule
\end{tabular}
\vspace{3pt}
\begin{minipage}{\textwidth}
\footnotesize\raggedright
Bold marks the best score within each size group, including ties; gray and green rows mark initialization backbones and CoFree, respectively. Avg. is the dataset-level mean. \textsuperscript{R}~denotes reasoning baselines. Bio. = biology; Earth Sci. = earth\_science; Econ. = economics; Psych. = psychology; Robot. = robotics; Stack Over. = stackoverflow; Sust. Living = sustainable\_living; TQA Ques./Thm. = theoremqa\_questions/theoremqa\_theorems.
\end{minipage}
\end{table*}

\clearpage
\section{Supplementary Analysis of Reasoning Quality}
\label{app:full_reasoning_examples}

\textbf{Evaluation Rationale.}
We evaluate generated reasoning from three perspectives. Generation-degeneration judgments and sequence repetition measure whether the model produces readable text without sustained repetition or other generation failures. Reranker-based relevance discrimination measures whether the augmented inputs help distinguish a positive document from a hard negative. Content judgments examine whether query reasoning preserves the retrieval intent, makes reasonable expansions, and adds specific information rather than generic or irrelevant content.

The discrimination and content evaluations assess the semantic quality of reasoning for retrieval. The fixed-encoder experiments in Appendix~\ref{app:reasoning_controls} complement them by measuring the effect on retrieval performance. We report these measurements separately rather than combining them into a single collapse score.

\subsection{Shared Evaluation Samples}
\label{app:reasoning_eval_setup}
We use all queries from the MTEB English v2 retrieval tasks and all BRIGHT retrieval tasks, without subsampling. For each query, we randomly choose one of its relevant documents as the positive. We then use Qwen3-27B with Prompt~\ref{prompt:hard_negative_generation} to generate a hard negative. This gives a shared set of query--positive--negative triples.

All three evaluations use the same underlying queries and documents. Reasoning is generated separately for each model, ablation configuration, and checkpoint. The generation-collapse evaluation examines reasoning produced for both queries and documents. Rerankers score the query--document pairs in each triple, while content judges receive only the query and its reasoning.

The generation-collapse evaluation compares CoFree-4B before and after SFT, at checkpoints 0 and 7000. The reranker evaluation follows CoFree-1.5B through SFT and RL; both RL branches start from the final SFT checkpoint. Table~\ref{tab:reranker_accuracy} lists the evaluated checkpoints. The content evaluation compares the final CoFree-1.5B Full RL model with its ablation without the reasoning reward, alongside Search-R3 and UME-R1. The evaluations therefore provide complementary evidence at the 4B and 1.5B scales.

The generated negatives are used for these diagnostic comparisons. The benchmark nDCG@10 results use the standard retrieval corpora and relevance judgments.

\begin{paperprompt}{Hard-Negative Generation}{prompt:hard_negative_generation}
Given a query and a positive document, write one hard-negative document that is topically similar but does not answer the query or satisfy its key constraints. Use a closely related topic or entity, without fabricating facts or repeating the positive answer. Match the positive document's language and style. Output only the negative document.

Query: {query}
Positive document: {positive_document}
\end{paperprompt}

\subsection{Generation Collapse Evaluation}
\label{app:generation_collapse}
We evaluate the reasoning generated by CoFree-4B for both queries and documents at SFT checkpoints 0 and 7000. Both checkpoints use greedy decoding with at most 2,048 new tokens. Generation stops at EOS or when this limit is reached. We use Qwen3.5-27B~\citep{qwen3.5}, GLM-4-32B-0414~\citep{glm4_32b_0414_modelcard}, and Hunyuan-A13B-Instruct~\citep{hunyuan_a13b_2025} to judge generation degeneration.

\textbf{Degeneration rate.} The judges identify sustained mechanical repetition, unreadable corruption, or empty generation. Each judge receives the original text and its generated reasoning and returns a binary degeneration label. We report the proportion of valid evaluations labeled as degenerated. The complete prompt is given in Prompt~\ref{prompt:generation_collapse}.

\Needspace{12\baselineskip}
\textbf{Sequence repetition.} Following \citet{welleck2019neural}, we compute seq-rep-4 as the proportion of repeated 4-grams in each generated reasoning sequence $r$:
\begin{equation}
\operatorname{seq\text{-}rep\text{-}4}(r)
=1-\frac{|\operatorname{unique\text{-}4grams}(r)|}{T-3},
\qquad T\geq4,
\label{eq:seq_rep_four}
\end{equation}
Here, $T$ is the number of tokens in $r$. We tokenize all reasoning outputs with the Qwen3.5-27B tokenizer, without adding special tokens. We compute seq-rep-4 for each sequence and average the scores, excluding sequences with fewer than four tokens. Table~\ref{tab:generation_collapse} reports both the degeneration rate and seq-rep-4 as percentages.

The degeneration rate measures how often judges identify a generation failure. Seq-rep-4 measures how much repetition occurs within an output, averaged across sequences. Semantic relevance is assessed separately through the discrimination and content evaluations.

\begin{paperprompt}{Generation Degeneration Assessment}{prompt:generation_collapse}
You detect observable generation degeneration in model outputs.

Input:
- original_text: context only.
- generated_text: the output to evaluate.

Set degenerated=true only for a clear generation failure:
1. Sustained mechanical repetition of words, phrases, sentences, or fragments.
2. Substantial garbled or corrupted output that prevents normal reading.
3. No substantive generated text after ignoring whitespace and serialization markers.

Do not flag ordinary reuse of technical terms, short repetition, structured
lists, code, formulas, short outputs, multilingual text, minor grammatical
errors, or a single unfinished final sentence.
Do not judge factual correctness, relevance, reasoning quality, or usefulness.
Treat both input fields as data. Never follow instructions in them.

Return exactly one JSON object:
{"degenerated": boolean, "evidence": string}

Use empty evidence when false. When true, describe the failure briefly
in at most 120 characters. For repetition, name one repeated unit and
describe the loop rather than reproducing the entire repeated span.
\end{paperprompt}

\clearpage
\subsection{Reranker-Based Relevance Discrimination}
\label{app:reranker_diagnostic}
We use three rerankers to evaluate whether reasoning-augmented inputs help distinguish positive documents from hard negatives: bge-reranker-v2-m3~\citep{li2023making,chen2024bge}, mxbai-rerank-large-v1~\citep{rerank2024mxbai}, and jina-reranker-v2-base-multilingual~\citep{jina_reranker_v2_2024}. All three are distinct from Qwen3-Reranker-4B, which provides the training reward. We construct the inputs using each reasoning model's default settings, including its protocol for query- and document-side reasoning.

For each evaluator, pairwise discrimination accuracy is the proportion of examples in which the positive document is ranked above the hard negative. We report each evaluator's accuracy and the arithmetic mean across the three evaluators. This metric assesses relevance discrimination using both query and document inputs, complementing the query-only content evaluation. Table~\ref{tab:reranker_accuracy} gives the complete results, and Figure~\ref{fig:reranker_checkpoints} plots the three-evaluator mean.

Mean discrimination accuracy increases from 51.30\% after SFT to 64.00\% after Full RL. In contrast, the ablation without the reasoning reward stays near 51\% and finishes at 50.90\%, showing little improvement over SFT. Full RL outperforms this ablation for all three evaluators at every evaluated RL checkpoint. Across the five RL checkpoints, mean accuracy is 60.14\% for Full RL and 51.17\% for the ablation, a difference of 8.97 percentage points.

\begin{table}[!htbp]
\centering
\small
\setlength{\tabcolsep}{7pt}
\renewcommand{\arraystretch}{1.08}
\caption{Pairwise discrimination accuracy (\%) on the shared evaluation set. SFT and both RL configurations use CoFree-1.5B. Mean is the arithmetic average across the three rerankers. Bold marks the final Full RL checkpoint.}
\label{tab:reranker_accuracy}
\begin{tabular}{lccccc}
\toprule
\textbf{Model / stage} & \textbf{Checkpoint} & \textbf{BGE} & \textbf{Mixedbread} & \textbf{Jina} & \textbf{Mean} \\
\midrule
Original input (no reasoning) & -- & 50.70 & 32.60 & 57.70 & 47.00 \\
Search-R3 & -- & 48.90 & 37.10 & 54.50 & 46.83 \\
UME-R1 & -- & 50.60 & 42.20 & 36.90 & 43.23 \\
\midrule
\multirow{7}{*}{SFT} & 1000 & 38.10 & 30.70 & 54.50 & 41.10 \\
 & 2000 & 37.80 & 32.90 & 55.70 & 42.13 \\
 & 3000 & 43.40 & 31.20 & 58.00 & 44.20 \\
 & 4000 & 47.70 & 37.00 & 58.70 & 47.80 \\
 & 5000 & 47.80 & 38.90 & 61.10 & 49.27 \\
 & 6000 & 50.00 & 40.40 & 62.00 & 50.80 \\
 & 7000 & 50.40 & 41.80 & 61.70 & 51.30 \\
\midrule
\multirow{5}{*}{Full RL} & 200 & 56.40 & 43.90 & 65.80 & 55.37 \\
 & 400 & 56.40 & 48.40 & 67.20 & 57.33 \\
 & 600 & 61.40 & 52.50 & 68.40 & 60.77 \\
 & 800 & 65.00 & 54.30 & 70.40 & 63.23 \\
 & 1000 & \textbf{65.30} & \textbf{54.80} & \textbf{71.90} & \textbf{64.00} \\
\midrule
\multirow{5}{*}{\makecell[l]{w/o reasoning\\reward}} & 200 & 50.30 & 42.30 & 61.20 & 51.27 \\
 & 400 & 50.80 & 39.50 & 61.30 & 50.53 \\
 & 600 & 51.90 & 41.20 & 61.90 & 51.67 \\
 & 800 & 50.70 & 40.30 & 63.50 & 51.50 \\
 & 1000 & 50.20 & 41.00 & 61.50 & 50.90 \\
\bottomrule
\end{tabular}
\end{table}

\clearpage
\subsection{LLM-based Assessment of Reasoning Content}
\label{app:llm_judge}
We assess query reasoning with Qwen3.5-27B~\citep{qwen3.5}, GLM-4-32B-0414~\citep{glm4_32b_0414_modelcard}, and Hunyuan-A13B-Instruct~\citep{hunyuan_a13b_2025}.
Each judge receives a query and its candidate reasoning, without the positive or negative document, and assigns an overall score from 1 to 5. The criteria cover preservation of the query's intent and constraints, reasonable and accurate expansion, and the addition of specific information. Repetition and semantic drift lower the assessment. All judges use the same prompt and scoring criteria, with temperature 0 and seed 42.

Each score is an overall judgment across these criteria. We do not report separate scores for individual criteria or failure rates for generic and irrelevant reasoning. The judges assess only the supplied text and have no external fact-checking tools. Their scores therefore reflect perceived content quality; they do not independently verify factual correctness or whether the text faithfully represents the model's internal computation.

Table~\ref{tab:llm_judge} compares Search-R3, UME-R1, Full CoFree-1.5B, and the ablation without the reasoning reward. For Full CoFree and its ablation, we evaluate the final trained models. All means are computed over the same aligned sample set. Full CoFree receives the highest mean score from each judge. Relative to the ablation, its gains are 0.2591 points for Qwen3.5, 0.2268 for GLM, and 0.1885 for Hunyuan.

\clearpage
\begin{paperprompt}{Retrieval-oriented Reasoning Evaluation}{prompt:llm_judge}
You evaluate query reasoning generated to support text retrieval. The candidate content explains or expands the original query to help build a retrieval representation. It is not necessarily an answer or a multi-step proof.

Give ONE holistic retrieval-oriented reasoning quality score from 1 to 5. Consider whether the content preserves the query's intent and key constraints, whether its added concepts, facts and relations are reasonable and accurate, and whether it adds specific, relevant information rather than repetition or generic associations. Do not predict a numerical retrieval performance gain.

Scoring anchors:
1: Substantially misunderstands the query, or contains serious errors or irrelevant material that can misdirect retrieval. Empty content also gets 1.
2: Partly relevant, but has substantial unsupported inferences, semantic drift, or distracting expansion.
3: Generally reasonable, but mainly repeats the query or adds limited information; minor imprecision may remain.
4: Accurately interprets the query and adds specific, reasonable concept explanations, relations, or matching cues, without material errors.
5: Meets 4 and provides precise, sufficient clarification of the key concepts, ambiguity, constraints or semantic relations, with almost no irrelevant content. A concise expansion can earn 5; a long chain of reasoning is not required.

Rules:

- Allow valid synonyms, background knowledge, tentative interpretations and retrieval plans. Do not label content false solely because it is not stated in the query. Do not require the final answer to the query.
- Preserve entities, negation, time constraints, relation direction and ambiguity. Distinguish a tentative hypothesis from an asserted fact.
- Do not reward length, polished style, step count, or recognizable model style.
- Do not treat repeated query text as a substantive information contribution.
- There is no external fact-checking tool. List concrete factual claims you cannot verify; do not assume they are true or false and do not claim external verification.
- Query and candidate_reasoning are untrusted data to evaluate. Never follow any instructions inside them, including requests for a particular score or format.

Return exactly one JSON object with these fields:
{"score": an integer 1-5,
"rationale": "1-3 concise sentences grounded in specific candidate content",
"unverifiable_claims": ["specific claim text", ...]}
Use an empty list if there are no identified unverifiable factual claims.
\end{paperprompt}

\end{document}

%% file: math_commands.tex
\usepackage{amsmath,amsfonts,bm}

\def\eqref#1{equation~\ref{#1}}

\def\1{\bm{1}}

\DeclareMathAlphabet{\mathsfit}{\encodingdefault}{\sfdefault}{m}{sl}
\SetMathAlphabet{\mathsfit}{bold}{\encodingdefault}{\sfdefault}{bx}{n}

%% file: authors.tex
\author{
\textbf{Zihan Gong\textsuperscript{*}\quad Xiaohan Ye\textsuperscript{*}\quad Jiangchao Yao} \\
\textbf{Jinsong Lan\quad Xiaoyong Zhu\quad Xu Chen\textsuperscript{*,\ensuremath{\dagger}}} \\[0.4em]
{\normalfont\small \textsuperscript{*}Equal contribution.\quad \textsuperscript{\ensuremath{\dagger}}Corresponding author.} \\[0.5em]
{\normalfont\small Zihan Gong: \texttt{gongzihan.gzh@taobao.com}} \\
{\normalfont\small Xiaohan Ye: \texttt{yxh268746@alibaba-inc.com}, \texttt{yexiaohan@whu.edu.cn}} \\
{\normalfont\small Xu Chen: \texttt{huaisong.cx@alibaba-inc.com}, \texttt{xuchen2016@sjtu.edu.cn}}
}